\documentclass[manuscript]{aastex631}

\usepackage{threeparttable}

\begin{document}

%\title{Performance Simulations of Exoplanet Transmission Spectrum Observations with CSST}
\title{The capability of CSST in characterizing planetary atmospheres. I. transmission spectroscopy of hot Jupiters}

\author[0009-0001-5127-8577]{Zibo Liu}
\affiliation{National Astronomical Observatories, Chinese Academy of Sciences,\\
Beijing 100101, PR China}

\author[0000-0002-9702-4441]{Wei Wang}
\affiliation{National Astronomical Observatories, Chinese Academy of Sciences,\\
Beijing 100101, PR China}
\affiliation{School of Astronomy and Space Science, University of Chinese Academy of Sciences, Beijing 100049, PR China}

\author[0000-0003-1207-3787]{Meng Zhai}
\affiliation{National Astronomical Observatories, Chinese Academy of Sciences,\\
Beijing 100101, PR China}

\author[0009-0005-8304-4192]{Jinpeng Wang}
\affiliation{National Astronomical Observatories, Chinese Academy of Sciences,\\
Beijing 100101, PR China}
\affiliation{School of Astronomy and Space Science, University of Chinese Academy of Sciences, Beijing 100049, PR China}

\author[0009-0001-1490-2991]{Qinglin Ouyang}
\affiliation{Department of Astronomy, University of Science and Technology of China, Hefei 230026, PR China}

\author[0000-0001-9585-9034]{Fei Yan}
\affiliation{Department of Astronomy, University of Science and Technology of China,
Hefei 230026, PR China}
\affiliation{Deep Space Exploration Laboratory, Hefei, Anhui 230026, PR China}

\author[0000-0003-0740-5433]{Guo Chen}
\affiliation{CAS Key Laboratory of Planetary Science, Purple Mountain Observatory, Chinese Academy of Sciences,\\
Nanjing 210023, PR China}

\author[0000-0002-0483-0445]{Dongdong Ni}
\affiliation{Institute of Science and Technology for Deep Space Exploration, Nanjing University,\\ Suzhou 215163, PR China}
\affiliation{State Key Laboratory of Lunar and Planetary Sciences, Macau University of Science and Technology,\\ Macau, PR China}

\author[0000-0002-2841-047X]{Yong Zhao}
\affiliation{State Key Laboratory of Lunar and Planetary Sciences, Macau University of Science and Technology,\\ Macau, PR China}

\author[0009-0008-3430-1027]{Yujuan Liu}
\affiliation{National Astronomical Observatories, Chinese Academy of Sciences,\\
Beijing 100101, PR China}

\author[0000-0003-4276-1767]{Fei Zhao}
\affiliation{National Astronomical Observatories, Chinese Academy of Sciences,\\
Beijing 100101, PR China}

\author[0000-0002-8980-945X]{Gang Zhao}
\affiliation{National Astronomical Observatories, Chinese Academy of Sciences,\\
Beijing 100101, PR China}
\affiliation{Institute of Space Sciences, Shandong University,\\
Shandong, China}
%\affiliation{School of Astronomy and Space Science, University of Chinese Academy of Sciences, Beijing 100049, PR China}
\correspondingauthor{Wei Wang}
\email{wangw@nao.cas.cn}

%% Note that the \and command from previous versions of AASTeX is now
%% depreciated in this version as it is no longer necessary. AASTeX 
%% automatically takes care of all commas and "and"s between authors names.

%% AASTeX 6.31 has the new \collaboration and \nocollaboration commands to
%% provide the collaboration status of a group of authors. These commands 
%% can be used either before or after the list of corresponding authors. The
%% argument for \collaboration is the collaboration identifier. Authors are
%% encouraged to surround collaboration identifiers with ()s. The 
%% \nocollaboration command takes no argument and exists to indicate that
%% the nearby authors are not part of surrounding collaborations.

%% Mark off the abstract in the ``abstract'' environment. 
\begin{abstract}

Transmission spectroscopy has become a primary tool for probing exoplanetary atmospheres, enabling constraints on their chemical compositions and providing limited information on their thermal properties. We assess the potential of the upcoming Chinese Space Station Telescope (CSST) for exoplanet atmospheric characterization through transmission spectroscopy. Theoretical spectra of hot gas planets are generated and used to simulate slitless spectroscopic observations with the CSST across the ultraviolet-to-near-infrared range. Atmospheric retrievals performed on the simulated data are compared with the input models to assess the robustness and accuracy of parameter determinations. We find that multi-band observations across three wavelength channels, each with two transits can place meaningful constraints on key atmospheric parameters. For multi-band observations that account for correlated (red) noise, future CSST observations are expected to achieve constraints that are comparable to, or in some cases slightly weaker than, those of the Hubble Space Telescope (HST), depending on the noise level and observing strategy. We conclude that CSST will provide unique and complementary constraints on the chemical compositions and physical properties of exoplanetary atmospheres, particularly for atomic species, metal-bearing molecules, and scattering processes accessible in the UV and optical, thereby complementing JWST's infrared sensitivity to molecular species.

\end{abstract}

%% Keywords should appear after the \end{abstract} command. 
%% The AAS Journals now uses Unified Astronomy Thesaurus concepts:
%% https://astrothesaurus.org
%% You will be asked to selected these concepts during the submission process
%% but this old "keyword" functionality is maintained in case authors want
%% to include these concepts in their preprints.

%% From the front matter, we move on to the body of the paper.
%% Sections are demarcated by \section and \subsection, respectively.
%% Observe the use of the LaTeX \label
%% command after the \subsection to give a symbolic KEY to the
%% subsection for cross-referencing in a \ref command.
%% You can use LaTeX's \ref and \label commands to keep track of
%% cross-references to sections, equations, tables, and figures.
%% That way, if you change the order of any elements, LaTeX will
%% automatically renumber them.
%%
%% We recommend that authors also use the natbib \citep
%% and \citet commands to identify citations.  The citations are
%% tied to the reference list via symbolic KEYs. The KEY corresponds
%% to the KEY in the \bibitem in the reference list below. 

\section{Introduction} \label{sec:intro}

A major goal of contemporary exoplanet science is to understand the physical and chemical diversity of planetary atmospheres and the processes that shape them. Transmission spectroscopy has become a primary tool for probing exoplanetary atmospheres, enabling measurements of chemical composition and limited constraints on their thermal properties through wavelength-dependent transit depths. The first atmospheric detection, sodium in HD 209458\,b \citep{Charbonneau2002}, was followed by numerous identifications of species, particularly Na and H$_2$O, in a variety of exoplanets \citep{Kreidberg2014,McCullough2014,Redfield2008,Nikolov2018}. 

Space-based observations are particularly powerful because they avoid telluric contamination and provide broad wavelength coverage. The Hubble Space Telescope (HST) has played a central role in ultraviolet (UV) and visible (VIS), and near-infrared (NIR) wavelengths. In particular, WFC3 observations have provided numerous measurements of water vapor absorption near 1.4 $\mu$m \citep[e.g.,][]{Kreidberg2014}, while UV/VIS observations have enabled detections of short-wavelength absorbers such as SiO~\citep{Lothringer2022}. More recently, the James Webb Space Telescope (JWST) has delivered unprecedented infrared constraints on atmospheric chemistry \citep{Ahrer2023,Rustamkulov2023}. However, JWST provides limited access to the optical regime (down to $\sim$0.6 $\mu$m) and lacks coverage in UV, while HST is approaching the end of its operational lifetime. Continued space-based observations in the UV-VIS bands are therefore crucial for constraining continuum slopes, alkali line cores, and high-altitude absorbers.

The Chinese Space Station Telescope (CSST) is a 2-meter space telescope planned for low Earth orbit, with an orbital period of approximately 95 minutes. Its off-axis optical design is expected to provide image quality comparable to that of HST \citep{Zhan2021}. Equipped with slitless spectroscopic capability provided by the Multi-band Imaging and Slitless Spectroscopy Survey Camera (MC), CSST will enable observations in three contiguous bands (GU: 255–400\,nm; GV: 400–620\,nm; GI: 620–1000\,nm) at $R \sim 200$ ~\citep{Gong2026}. This configuration will provide continuous UV--optical coverage extending to $\sim$1\,$\mu$m. Although primarily designed for wide-field surveys, its spectroscopic mode will enable time-series observations suitable for exoplanet transmission spectroscopy.

To optimize CSST's performance in exoplanet atmospheric studies, comprehensive simulations of its observational capabilities are needed prior to launch. This paper investigates CSST's potential for atmospheric characterization of hot Jupiters (HJs) and ultra-hot Jupiters (UHJs) via transmission spectroscopy, and we are meanwhile conducting similar simulations for Neptune-sized planets. We generate theoretical spectra for a representative sample of hot gas giants, simulate CSST observations, and perform atmospheric retrievals on the synthetic data. By comparing retrieved parameters with input models, we quantify the expected precision and limitations of future CSST observations in constraining atmospheric temperature, metallicity, and key chemical species.

The paper is organized as follows. Section \ref{sec:method} describes the methodology for simulating CSST observations and performing atmospheric retrievals. Section \ref{sec:results} presents the simulated spectra and the corresponding retrieval outcomes. Section \ref{sec:discussion} discusses the implications of these results, the impact of correlated noise, and observational strategies. Finally, Section \ref{sec:conclusion} summarizes the main conclusions and outlines prospects for future CSST exoplanet observations.

\section{Methods} \label{sec:method}

\subsection{Target sample selection and system parameters} \label{sec:method-target}

This study aims to evaluate CSST's capability to probe exoplanetary atmospheres via transmission spectroscopy. In this work, we focus only on hot Jupiters, for which large atmospheric scale heights are expected to produce detectable spectral features. As a baseline for the simulations, planetary atmospheres are assumed to be isothermal and in chemical equilibrium, allowing us to isolate the instrumental performance of CSST from additional atmospheric complexities.

To guide the selection of representative targets and to compare their relative observability, we employ a modified Transmission Spectroscopy Metric (TSM; \citealp{Kempton2018}), defined as

\begin{equation}
    \mathrm{TSM} = (\mathrm{Scale\ factor}) \times \frac{R_{\rm p}^3 T_{\mathrm{eq}}}{M_{\rm p} R_{\rm s}^2} \times 10^{-m_V/5},
\end{equation}

where $R_{\rm p}$ and $M_{\rm p}$ are the planetary radius and mass in Earth units, $R_{\rm s}$ is the stellar radius in solar units, $T_{\mathrm{eq}}$ is the equilibrium temperature, and $m_V$ is the apparent stellar magnitude in $V$-band. Although the original TSM uses the $J$-band magnitude, we adopt the $V$-band magnitude here because CSST transmission spectroscopy in this work covers the near-UV and optical wavelengths, for which the optical brightness of the host star is more directly relevant. The scale factor is set to unity in this work, which is a reasonable approximation, given that all target planets are HJs.

\begin{table*}[htbp]
 \caption{Fundamental stellar parameters of the host stars for all target systems considered in this work.}
    \label{tab:star}
   % \textbf{Stellar parameters}
    \centering
    \begin{tabular}{cccccccl}
    \hline
        \noalign{\smallskip}
    Star name & $R_{\mathrm{s}}$ $(R_{\odot})$ & $T_{\mathrm{eff}}$(K) & log($g$)& [Fe/H] & $m_V$ & Distance ($pc$) & Reference \\
        \noalign{\smallskip}
    \hline
        \noalign{\smallskip}
    WASP-19 & 1.004 & 5568 & 4.45 & 0.15 & 12.248 & 268.325 & \cite{Wong2016}\\
    HAT-P-18 & 0.749 & 4803 & 4.57 & 0.1 & 12.597 & 161.4 & \cite{Hartman2011}\\
    WASP-96 & 1.05 & 5540 & 4.42 & 0.14 & 12.569 & 352.464 & \cite{Hellier2014}\\
    WASP-52 & 0.79 & 5000 & 4.582 & 0.03 & 12.192 & 174.818 & \cite{Hebrard2013}\\
    WASP-69 & 0.813 & 4715 & 4.535 & 0.144 & 9.873 & 49.9605 & \cite{Anderson2014}\\
    WASP-103 & 1.436 & 6110 & 4.22 & 0.06 & 12.402 & 375.647 & \cite{Gillon2014}\\
    WASP-178 & 1.67 & 9360 & 4.31 & 0.21 & 9.946 & 427.678 & \cite{Hellier2019}\\
        \noalign{\smallskip}
    \hline
    %星等
    \end{tabular}

\end{table*}

\begin{table*}[htbp]
    %\textbf{Planetary parameters}
    \caption{Physical and orbital parameters of the simulated planets, sorted by their atmospheric metallicity.}
    \label{tab:planet}
    \centering
    \footnotesize
    \begin{tabular}{ccccccccccccc}
    \hline
        \noalign{\smallskip}
    Planet Name & $R_{\mathrm{p}}$ & $M_{\mathrm{p}}$ & $\rho$ & $T_{\mathrm{eq}}$ & $i$ & $e$ & Period & $a/R_{\mathrm{s}}$ & [Fe/H] & C/O & $T_{\mathrm{exp}}$ & TSM\\
        \noalign{\smallskip}
    \hline
        \noalign{\smallskip}
     & ($R_{\mathrm{jup}}$) & ($M_{\mathrm{jup}}$) & ($g/cm^3$) & (K) & ($^{\circ}$) & & (day) & & & & (s) & \\
         \noalign{\smallskip}
    \hline
        \noalign{\smallskip}
    WASP-19\,b & 1.392 & 1.069 & 0.492 & 2120 & 78.78 & 0.002 & 0.788838989 & 3.46 & 0.95 & 0.47 & 200 & 83.5\\
    HAT-P-18\,b & 0.995 & 0.197 & 0.248 & 852 & 88.8 & 0.084 & 5.508023 & 16.04 & 0.84 & 1.05 & 250 & 101.8\\
    WASP-96\,b & 1.2 & 0.48 & 0.344 & 1285 & 85.6 & 0 & 3.4252602 & 9.28 & 0.53 & 0.78 & 350 & 57.0\\
    WASP-52\,b & 1.27 & 0.46 & 0.278 & 1315 & 85.35 & 0 & 1.7497798 & 7.38 & 0.33 & 0.41 & 200 & 151.5\\
    WASP-69\,b & 1.057 & 0.26 & 0.273 & 963 & 86.71 & 0 & 3.8681382 & 11.97 & -0.13 & 0.42 & 20 & 310.9\\
    WASP-103\,b & 1.528 & 1.49 & 0.518 & 2508 & 86.3 & 0 & 0.925542 & 2.978 & -0.3 & 1.03 & 300 & 42.7\\
    WASP-178\,b & 1.87 & 1.66 & 0.295 & 2470 & 84.41 & 0 & 3.3448285 & 6.588 & - & - & 40 & 169.2\\
    \noalign{\smallskip}
    \hline
    %曝光时间
    \noalign{\smallskip}
    \noalign{\smallskip}
    \end{tabular}
    \begin{tablenotes}
    \small
      \item $^a$ All parameters for each planetary system come from the same reference as the corresponding host star in Table~\ref{tab:star}, except for WASP-178\,b. $T_{\mathrm{exp}}$ refers to the assumed exposure time used in this work.
      \item $^b$ All $T_{\mathrm{eq}}$ correspond to the same assumption of zero Bond albedo and full heat redistribution.
      \item $^c$ For WASP-178\,b, $R_{\rm p}$, $i$ and $a/R_{\mathrm{s}}$ are taken from \citet{Lothringer2022}   
    \end{tablenotes}

\end{table*}

We select six gas giant exoplanets spanning a range of equilibrium temperatures, atmospheric metallicities, and C/O ratios. They all have high TSM values, and with basic stellar, planetary, and atmospheric parameters already well measured, as listed in Tables \ref{tab:star} and \ref{tab:planet}, among which $i$ and $e$ denote the planet's orbital inclination and eccentricity, respectively. Atmospheric metallicities and C/O ratios are taken from HST retrieval results~\citep{Edwards2023} and are used as inputs for the forward models described in Section~\ref{sec:method-spectra}. 

In addition to the six planets described above, we include a seventh target that is used exclusively in Section \ref{sec:W178} for assessing the effects of time-correlated noise. For this system, we adopt atmospheric abundances inferred from free-chemistry retrievals of HST/UVIS transmission spectra, enabling a direct comparison between CSST simulations and existing HST optical observations.

\subsection{Preparation for generating theoretical transmission spectra}
\label{sec:method-spectra}

\begin{figure*}[htbp]
    \centering
    \fig{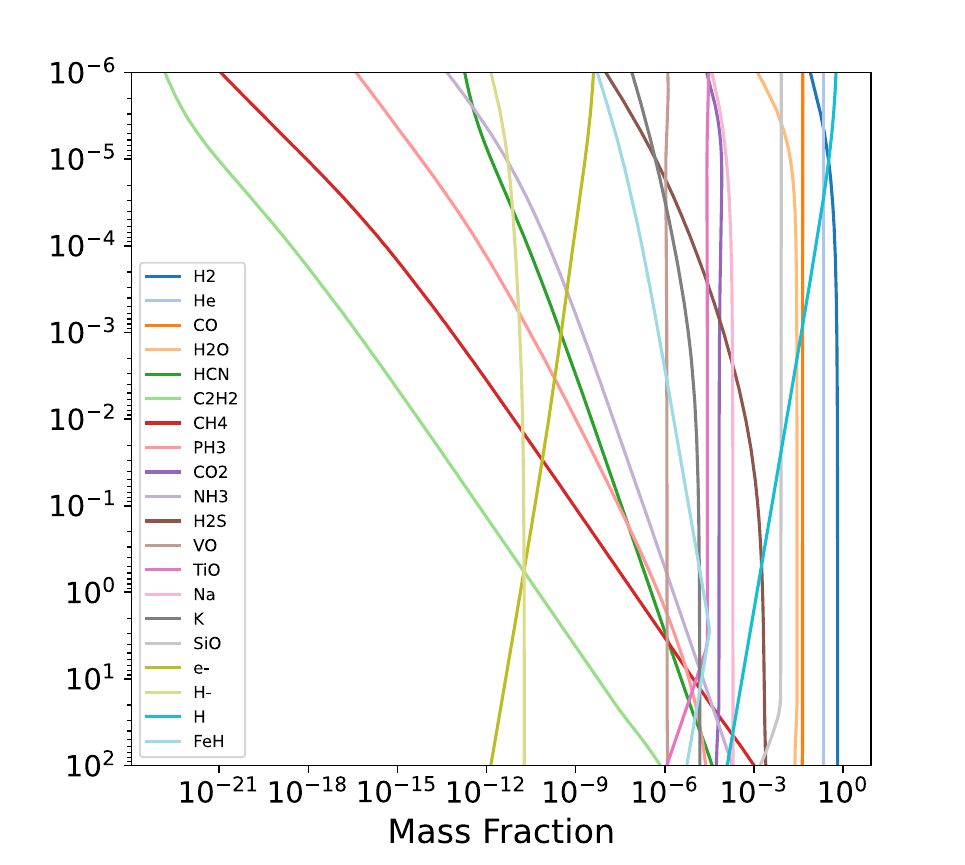}{0.5\textwidth}{}
    \caption{Mass fraction of chemical species in the atmosphere of WASP-19\,b as functions of pressure.}
    \label{fig:p-t-comp}
\end{figure*}

To generate a theoretical transmission spectrum for a particular planet, 3 procedures are to be conducted in advance, as listed below.
%\begin{enumerate}
\begin{itemize}
     \item  To generate, for each host star, a theoretical stellar spectrum $F_{\rm s}(\lambda)$ that has not been transmitted through planetary atmospheres;
     \item To generate a pure planet transmission spectrum $F_{\rm p, trans}(\lambda)$ for the corresponding planet.
     \item Given the transit parameters,e.g., $R_{\rm p}/R_{\rm s}$, a transit light curve $F_{\rm lc}(t)$ is produced, which is to be used to scale the fraction of planet atmospheric transmission as compared to stellar radiation. Limb darkening is not included, as it is expected to have a secondary effect on the transmission spectrum. A more detailed treatment including limb darkening will be explored in future work.

\end{itemize}
%\end{enumerate}
 
For the first step, stellar spectra are derived from the PHOENIX high-resolution (R$\geq$ 25000) stellar atmosphere models~\citep{Husser2013}. The PHOENIX grid spans effective temperatures $T_{\rm eff}$ from 2300 to 7000\,K in steps of 100\,K, surface gravities $\log g$ from 0.0 to 6.0 in steps of 0.5\,dex, and metallicities [Fe/H] from $-2.0$ to $+1.0$ in steps of 0.5\,dex. For each target star, we select the synthetic spectrum with the closest values of $T_{\rm eff}$, $\log g$, and [Fe/H].

Secondly, theoretical transmission spectra of exoplanetary atmospheres with resolution of 1000 are generated using \texttt{petitRADTRANS} \citep[hereafter pRT,][]{Molliere2019}. The atmospheres are assumed to be isothermal and in chemical equilibrium, with the atmospheric temperature set equal to the planetary equilibrium temperature. We assume a cloud-free atmosphere in the forward modeling to isolate the spectral features of gaseous species; the inclusion of clouds and hazes will be explored in future work. The atmospheric structure is defined on a pressure grid spanning from 100 to $10^{-6}$ bar with 200 layers. A reference pressure of 0.01 bar is adopted, at which the planetary radius is defined, and $\log g$ is taken from the parameters listed in Table~\ref{tab:planet}. Atmospheric metallicities and C/O ratios are adopted from the HST retrieval results reported by~\citet{Edwards2023}. Chemical equilibrium abundances are calculated internally using \texttt{easyCHEM}~\citep{Molliere2017}. The opacity sources included in the calculations comprise H$_2$, He, CO, H$_2$O, HCN, C$_2$H$_2$, CH$_4$, PH$_3$, NH$_3$, H$_2$S, VO, TiO, Na, K, SiO and FeH. In addition, Rayleigh scattering by H$_2$, He, CO, CO$_2$, CH$_4$, and H$_2$O, as well as collision-induced absorption (CIA) from H$_2$–H$_2$, H$_2$–He, CO$_2$–CO$_2$, and H$_2$O–H$_2$O, are taken into account. For the UHJs WASP-103\,b ($T_{\rm eq} = 2508$ K) and WASP-178\,b ($T_{\rm eq} = 2470$ K), we additionally include the contribution of H$^-$ opacity to the continuum for both theoretical spectrum and retrieved spectrum.

\begin{figure}
    \centering
    \gridline{
    \fig{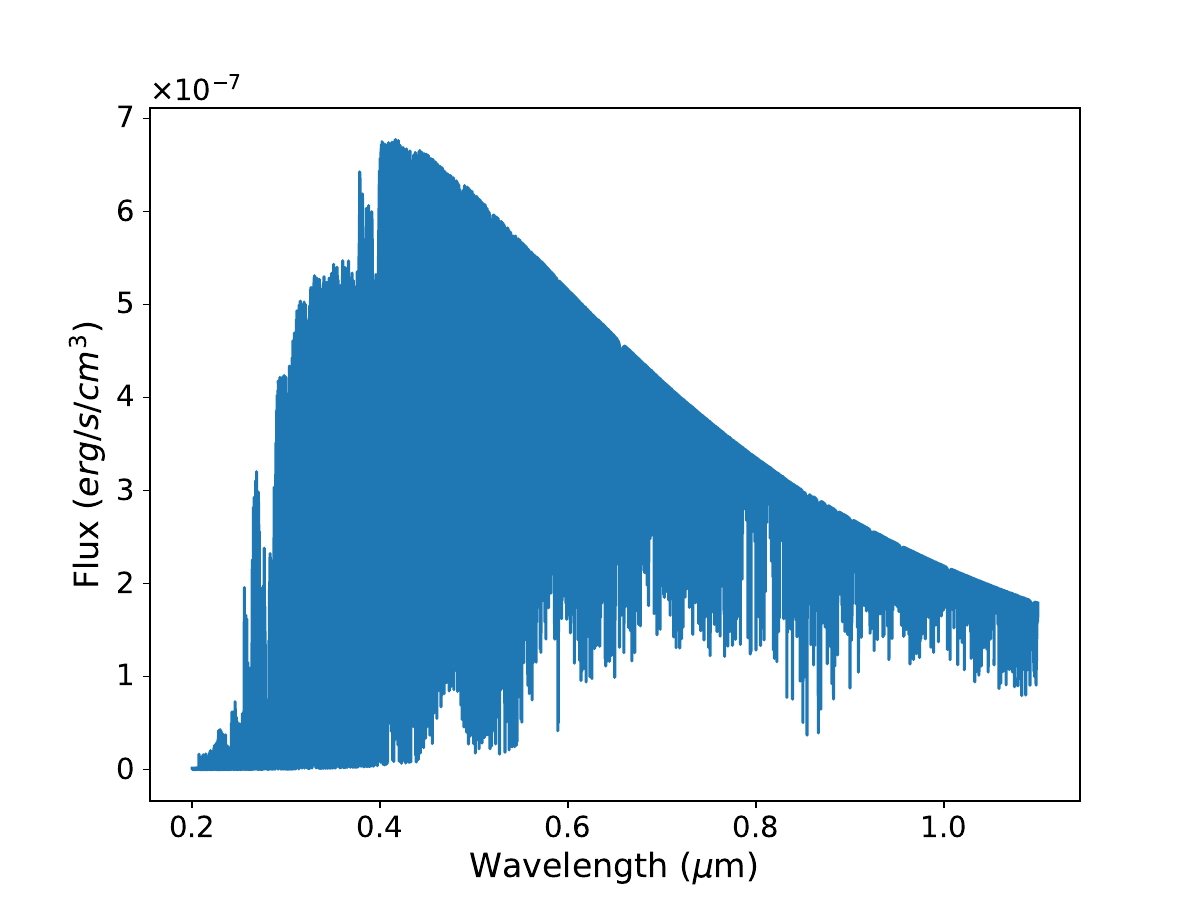}{0.45\textwidth}
    {(a) WASP-19}
    \fig{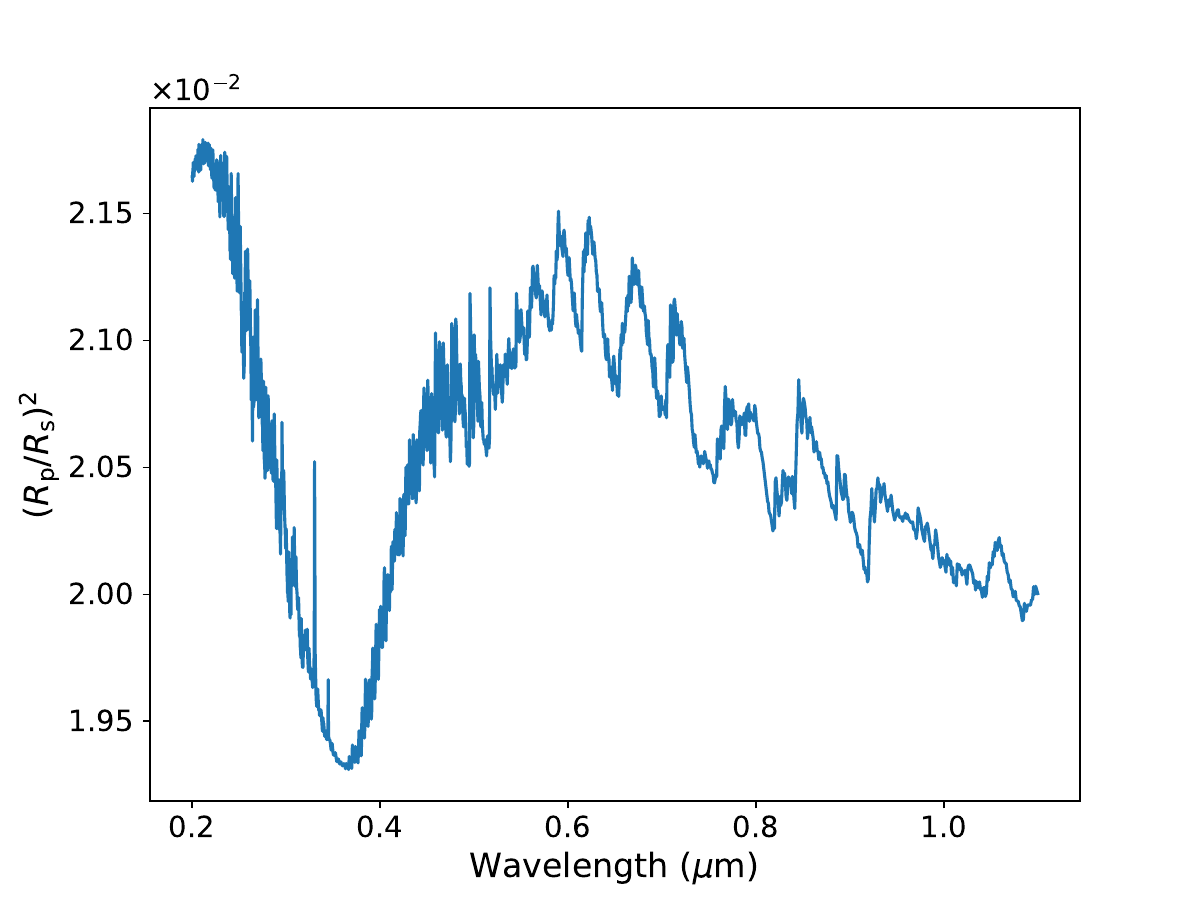}{0.45\textwidth}
    {(b) WASP-19\,b}
    }
    \caption{Theoretical spectrum of WASP-19 generated by PHOENIX model (\texttt{Left}, $R \geq 25000$), and WASP-19\,b generated by \texttt{petitRADTRANS} (\texttt{Right}, $R = 1000$).}
    \label{fig:theoretical_spectra}
\end{figure}

Then, for each star-planet system, we use the \texttt{batman} package~\citep{Kreidberg2015} to generate the model transit light curve, which is used to simulate the star-planet system spectrum for a given time $t$, i.e., $F(\lambda,t)=F_{\rm s}(\lambda)-F_{\rm p, trans}(\lambda)\times \frac{1-F_{\rm lc}(t)}{\mathrm{max}[1-F_{\rm lc}(t)]}$.

As an example, Fig.~\ref{fig:p-t-comp} illustrates the chemical abundance profiles we used in this work for WASP-19\,b, assuming $T_{\rm eq}=2120$ K and [Fe/H] $=0.95$. In addition, we show model stellar and planet spectra in Fig.~\ref{fig:theoretical_spectra} generated from the PHOENIX model and by \texttt{petitRADTRANS}, in Panels a and b, respectively. 

We note that \citet{Deming2017} showed that unresolved overlap between stellar and planetary absorption lines can introduce a resolution-linked bias in transmission spectroscopy, especially for planets orbiting late M dwarfs with dense molecular absorption bands. This effect is expected to be less severe for our targets, since the coolest host star in our sample has $T_{\mathrm{eff}}\simeq4700$ K. We also tested this effect by comparing spectra constructed with high-resolution stellar--planetary coupling and those obtained with our standard procedure, and found that the resulting differences are well below the typical instrumental noise level in our simulated spectra. We therefore retain the standard treatment in our simulations for computational efficiency. 

\begin{figure}
    \centering
    \includegraphics[width=0.9\linewidth]{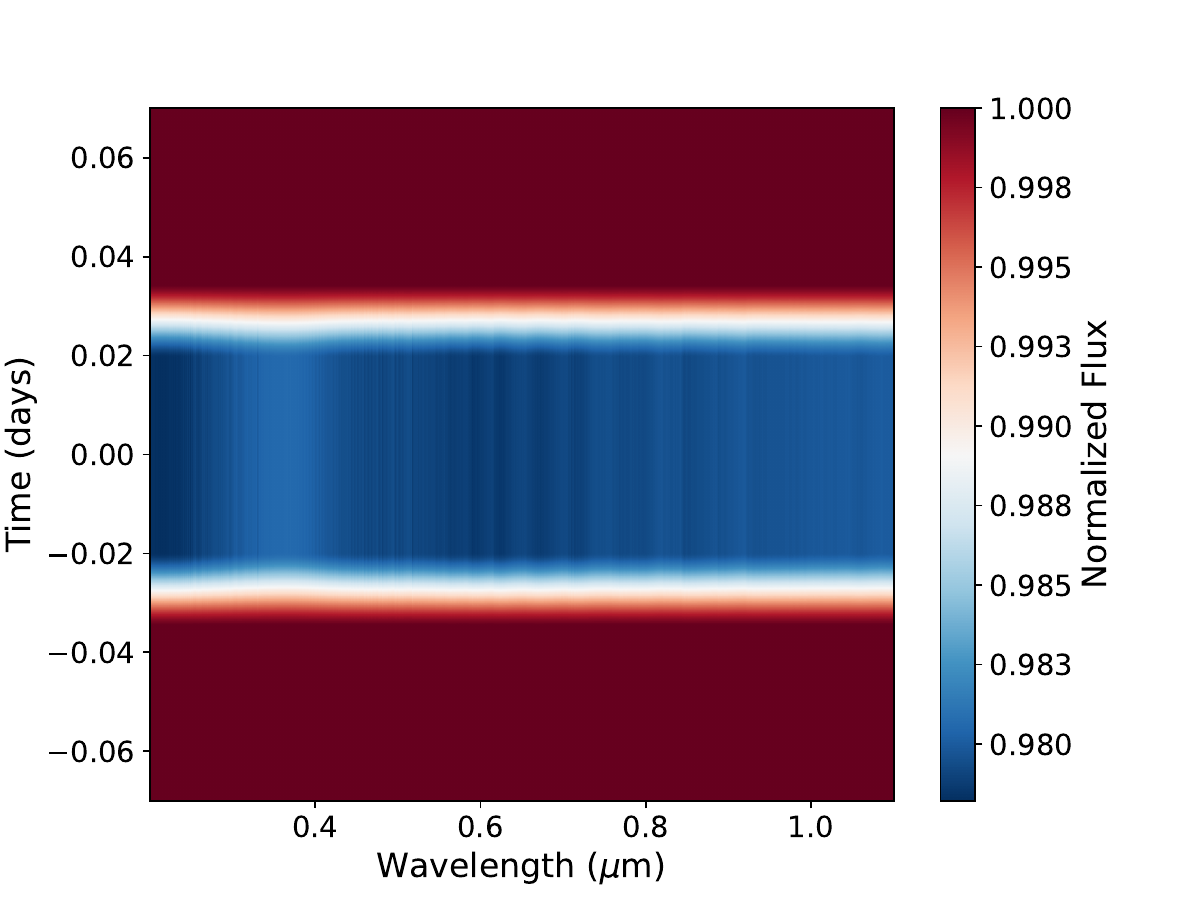}
    \caption{Theoretical 2D spectrum of WASP-19\,b varied with time and normalized by the stellar flux outside of transit.}
    \label{fig:2d_spectra}
\end{figure}

Finally, the time series of spectra $F(\lambda,t)$ covering one entire transit and some out-of-transit baseline is presented in Fig.~\ref{fig:2d_spectra} as a theoretical 2D spectrum, with a time interval between successive exposures consisting of the exposure time plus a fixed readout time of 40 seconds. In Fig.~\ref{fig:2d_spectra}, the baseline, ingress/egress, and the transit are distinguishable in the y-axis, while in the x-axis, spectral features from planet atmospheric absorption are apparent.

\subsection{Simulation of CSST transit observations} 
\label{sec:method-transit}

We adopt the CSST Multi-band Imaging and Slitless Spectroscopy Survey Camera (MC) configuration for all simulations in this work. The aforementioned model-generated one-dimensional (1D) spectrum $F(\lambda,t)$ is used as input to the CSST 1D slitless spectroscopic simulation package \texttt{SLS\_1D\_SPEC}\footnote{\url{https://csst-tb.bao.ac.cn/code/zhangxin/sls_1d_spec}} to produce a simulated 1D spectrum with simulated uncertainties. During the simulation, the detector readout noise is set to be 5\,e$^{-}$/pixel, the dark current is 0.02\,e$^{-}$/s/pixel, while the sky background levels are 0.019, 0.214, and 0.329\,e$^{-}$/s/pixel in the GU, GV and GI bands, respectively. The exposure time for each target in one spectrum is given in Table~\ref{tab:planet}, which is the maximum integration time before the CSST simulation software alerts to detector saturation.  Such a setup can thereby ensure the obtained spectra have the highest signal-to-noise ratio (SNR) achievable with CSST.  

\begin{figure}
    \centering
    \includegraphics[width=0.9\linewidth]{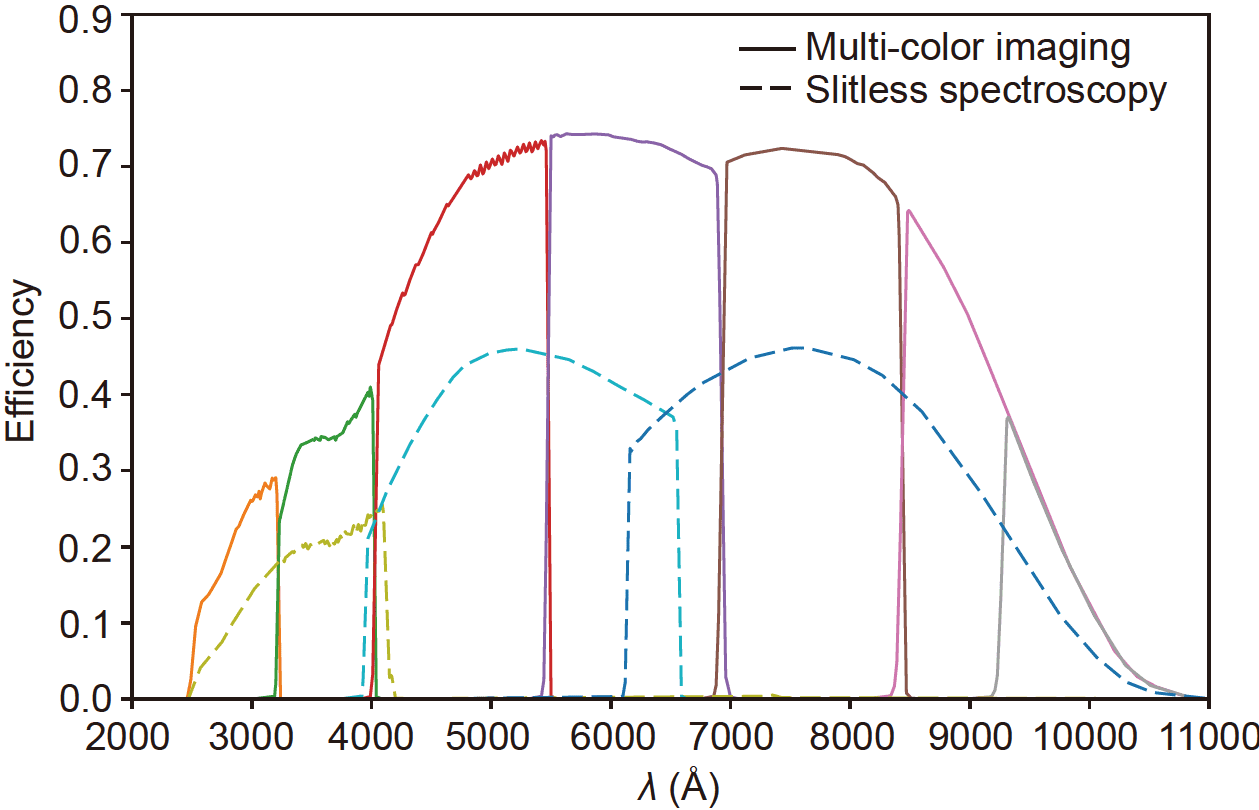}
    \caption{Total efficiency curves for the CSST slitless spectroscopic bands, adapted from Figure 7 of \citet{Zhan2021}. The total efficiency represents the system throughput calculated from the reflectivities of the telescope mirrors, the transmission of the survey-camera filters, the grating efficiencies, and the detector quantum efficiencies (QEs), including the wavelength-dependent CCD coatings adopted for the blue, visible, and red channels.}
    \label{fig:efficiency}
\end{figure}

CSST is planned to operate in a low-Earth orbit at an altitude of $\sim$400\,km, corresponding to an orbital period of $\sim$90 minutes. Due to Earth's occultation and the technical shutdown in the SAA (South Atlantic Anomaly) region, only a fraction of each orbit will be available for scientific observations. We first assume an orbital period of 95 minutes, with a continuous observing window of 45 minutes per orbit, following HST operational experience and enabling a direct comparison with previous HST observations.

\begin{figure*}[htbp]
    \centering
    \fig{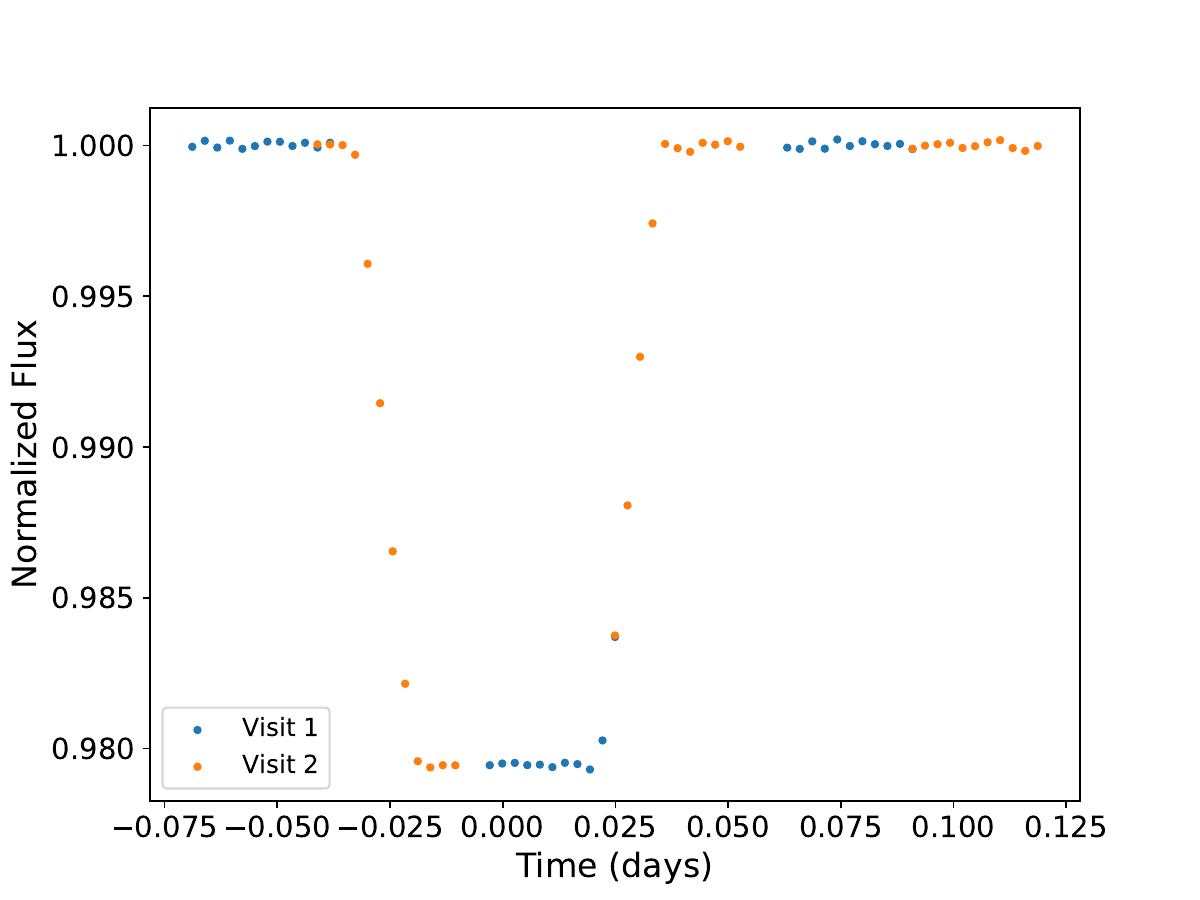}{0.85\textwidth}{}
    \caption{Simulated phase-folded transit white light curve of WASP-19\,b composed of 2 visits.}
    \label{fig:light-curve}
\end{figure*}

For most transiting exoplanets, transit duration is larger than 1\,hr; therefore, observations of at least two transits are required to obtain a complete transit light curve. Figure~\ref{fig:light-curve} presents the simulated phase-folded white-light curve of WASP-19\,b obtained in two visits, with markers indicating the middle times of individual exposures. By now, we have constructed the simulated observation data for transmission spectroscopy, i.e., $F^{\rm sim}(\lambda,t)$, together with their uncertainties, $E^{\rm sim}(\lambda,t)$. In this work, the light curves from different visits are combined directly for simplicity. In practice, they are typically fitted independently to account for visit-dependent systematics, which will be explored in future work.

\subsection{Spectral retrieval framework} \label{sec:method-retrieval}

Then we deal with the simulated time-series spectra $F^{\rm sim}(\lambda,t)$ following the classic method employed in ''real'' low-resolution transmission spectroscopy, as done for example in \citet{Ouyang2023,Ouyang2023b,Zhou2025}. In practice, from $F^{\rm sim}(\lambda,t)$, we first extract the white-light curve by integrating the flux over the full wavelength range in each band. The white-light curve is fitted using the affine-invariant Markov Chain Monte Carlo (MCMC) sampler \texttt{emcee}\citep{Foreman-Mackey2013} to constrain the broadband transit radius and other orbital parameters, and also an assessment of noise level and pattern in the light curve. The light curve extracted here do not include time-correlated (red) noise. The effect of red noise will be discussed in Section~\ref{sec:W178}.

We next construct spectroscopic light curves by binning the simulated spectra into wavelength channels of 20\,nm in the GU band and 10\,nm in the GV and GI bands, respectively. For each wavelength bin, the spectroscopic transit light curve is fitted again using MCMC, fixing all orbital parameters to the values derived from the white-light fit and allowing only the planetary radius to vary. This procedure yields the transit depths at each wavelength bin, and thus, the simulated transmission spectrum of the planet is produced.

Atmospheric retrievals are subsequently performed on the simulated transmission spectra using the nested sampling algorithm implemented in \texttt{petitRADTRANS}, based on the \texttt{MultiNest} framework~\citep{Buchner2014}. Two retrieval configurations are considered. In the first case, chemical equilibrium is assumed, with the planetary radius, atmospheric temperature, metallicity ([Fe/H]), and C/O ratio treated as free parameters. In the second case, a free-chemistry retrieval is adopted, retrieving the planetary radius, temperature, and the logarithm of the mass fractions of the dominant molecular species.

Uniform priors are adopted for radius, temperature, metallicity, and C/O ratio, while log-uniform priors are used for the mass fractions of individual chemical species. The prior ranges for radius, temperature, [Fe/H], C/O ratio, and log mass fractions are 0.8 to 2.0\,$R_{\mathrm{Jup}}$, 600 to 3000\,K, -1.0 to 2.0\,dex, 0.1 to 2.0, and -8.0 to -1.0\,dex, respectively.

\section{Results}\label{sec:results}

\subsection{Chemical equilibrium retrievals of atmospheric composition}\label{sec:chem-eq}

We first examine the retrieval performance under the assumption of chemical equilibrium (EQ). For each target, the simulated CSST transmission spectra cover two transits for each band by two visits. In this section, we summarize the overall recovery of key atmospheric parameters, while detailed comparisons for individual planets are deferred to Section\ref{sec:discussion}.

\begin{figure*}
\centering
\includegraphics[width=0.95\textwidth]{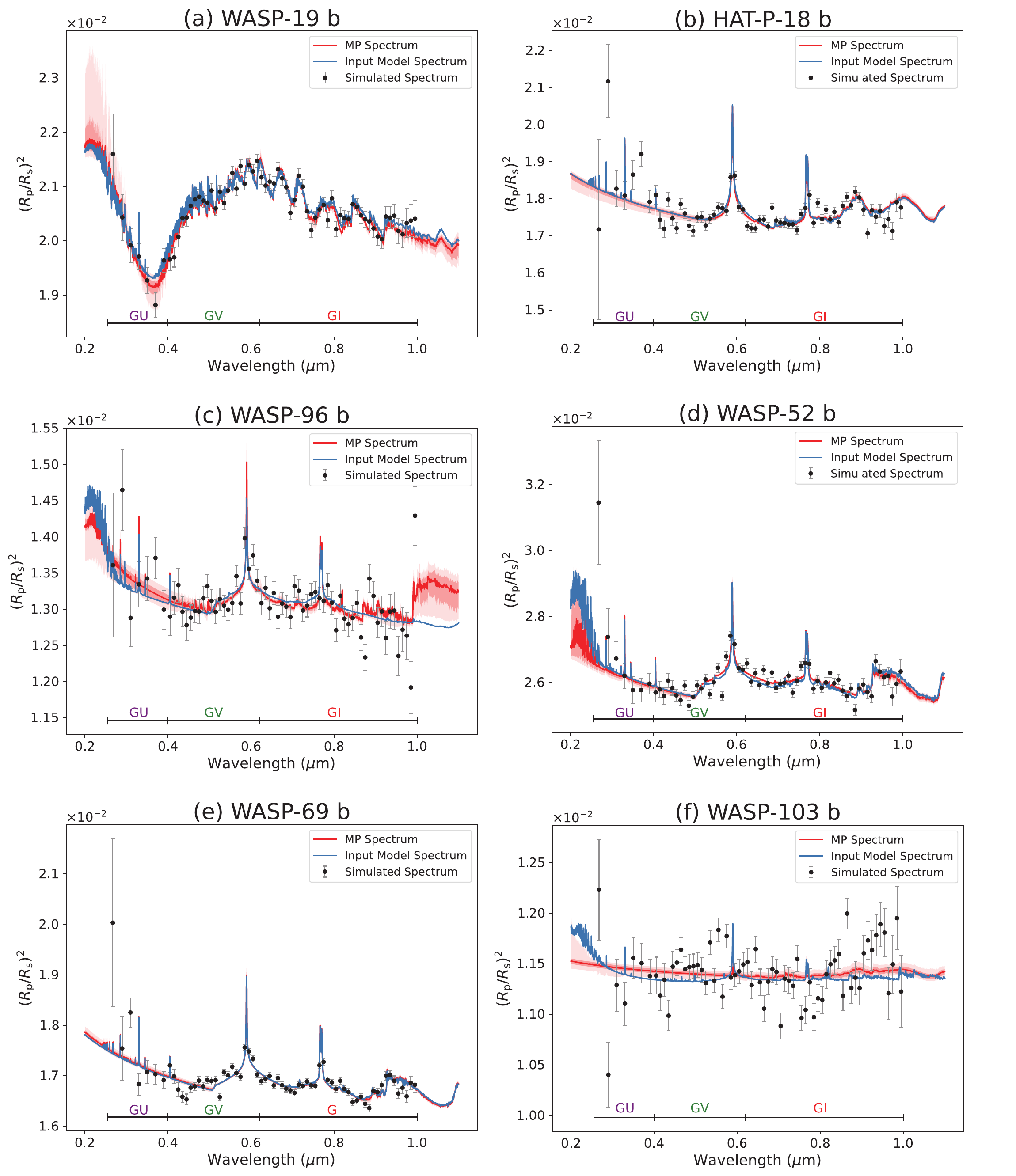}
\caption{The transmission spectra generated by pRT with input parameters (blue lines), deduced from the simulated CSST observation (black-filled circles with error bars) with 2 visits per band, and generated by pRT using the MPs (red lines), respectively. The shaded regions represent the 68\% and 95\% credible intervals of the posterior spectra.}
\label{fig:chem-eq-retrieval}
\end{figure*}

Figure~\ref{fig:chem-eq-retrieval} presents the transmission spectra generated directly by pRT, obtained from the CSST simulated observation, and the forward model spectra using the retrieved median posteriors (MPs). The retrieved planetary radius, atmospheric temperature, metallicity ([Fe/H]), and C/O ratio are summarized in Table~\ref{tab:chem-eq}, with the input values listed in parentheses.

\begin{table*}[htbp]
    %\textbf{Retrieved atmospheric parameters under chemical equilibrium}
  \caption{Retrieved atmospheric parameters for the EQ cases. The quoted uncertainties correspond to the 16th and 84th percentiles of the posterior distributions. Values in parentheses denote the input parameters listed in Table~\ref{tab:planet}.}
  %All retrievals are based on simulated observations covering two transits.
    \label{tab:chem-eq}
    \centering
    \begin{tabular*}{\textwidth}{@{\extracolsep{\fill}}ccccc}
    \hline
    \noalign{\smallskip}
    Planet Name & $R$($R_{\mathrm{jup}}$) & $T$(K) & [Fe/H] & C/O \\
    \noalign{\smallskip}
    \hline
    \noalign{\smallskip}
    WASP-19\,b & 1.387$^{+0.005}_{-0.008}$(1.392) & 2285$^{+133}_{-66}$(2120) & 1.00$^{+0.33}_{-0.12}$(0.95) & 0.28$^{+0.31}_{-0.52}$(0.47)\\
    HAT-P-18\,b & 0.995$^{+0.001}_{-0.001}$(0.995) & 844$^{+15}_{-10}$(852) & 0.91$^{+0.43}_{-0.20}$(0.84) & 1.05$^{+0.36}_{-0.19}$(1.05)\\
    WASP-96\,b & 1.207$^{+0.005}_{-0.005}$(1.200) & 1928$^{+190}_{-269}$(1285) & -0.11$^{+0.27}_{-0.26}$(0.53) & 1.56$^{+0.36}_{-0.39}$(0.78)\\
    WASP-52\,b & 1.269$^{+0.001}_{-0.002}$(1.270) & 1255$^{+47}_{-66}$(1315) & 0.77$^{+0.24}_{-0.15}$(0.33) & 0.57$^{+0.04}_{-0.04}$(0.41) \\
    WASP-69\,b & 1.058$^{+0.001}_{-0.001}$(1.057) & 990$^{+31}_{-30}$(963) & -0.25$^{+0.05}_{-0.05}$(-0.13) & 0.38$^{+0.10}_{-0.11}$(0.42)\\
    WASP-103\,b & 1.529$^{+0.001}_{-0.002}$(1.528) & 688$^{+689}_{-560}$(2508) & -0.33$^{+0.61}_{-0.64}$(-0.30) & 0.87$^{+0.76}_{-0.55}$(1.03)\\
    \noalign{\smallskip}
    \hline
    %曝光时间
    \end{tabular*}

\end{table*}

It is clear from Figure~\ref{fig:chem-eq-retrieval} that the EQ retrieval reproduces the overall spectral shape for all targets except WASP-103\,b, including the dominant optical and NIR absorption features across the CSST wavelength range. For most planets, the continuum level and major spectral features are well matched by the MP models.

Quantitatively, the input atmospheric parameters are recovered within $1\sigma$ uncertainties in the majority of cases. The agreement is particularly good for WASP-19\,b, WASP-52\,b, WASP-69\,b, and HAT-P-18\,b, where the retrieved values possess relatively smaller uncertainties, with no noticeable bias. Larger deviations are found for WASP-96\,b, where temperature and C/O tend to be overestimated and metallicity slightly underestimated. For WASP-103\,b in particular, the inclusion of H$^-$ opacity significantly suppresses molecular absorption features. As a consequence, both the simulated and retrieved spectra become nearly featureless within the CSST wavelength range, which limits the ability of the retrieval to constrain chemical abundances. These two targets also have comparatively lower TSM values, leading to weaker effective spectral signals and correspondingly broader parameter constraints.

Therefore, when performing EQ retrievals using the simulated CSST transmission spectra, the bulk atmospheric properties (Fe/H, C/O, etc.) of gas giants spanning a wide range of equilibrium temperatures can be recovered quite well in the adopted observational setup (one full transit covered by two visits), although for UHJs such as WASP-103\,b, the dominance of continuum opacity (e.g., H$^-$) can suppress spectral features and weaken compositional constraints. A more detailed analysis of parameter degeneracies and systematic trends is presented in Section \ref{sec:discussion} for each planet.

\subsection{Free-chemistry retrievals of atmospheric composition}
\label{sec:free-chem}

\begin{figure*}
\centering
\includegraphics[width=1\textwidth]{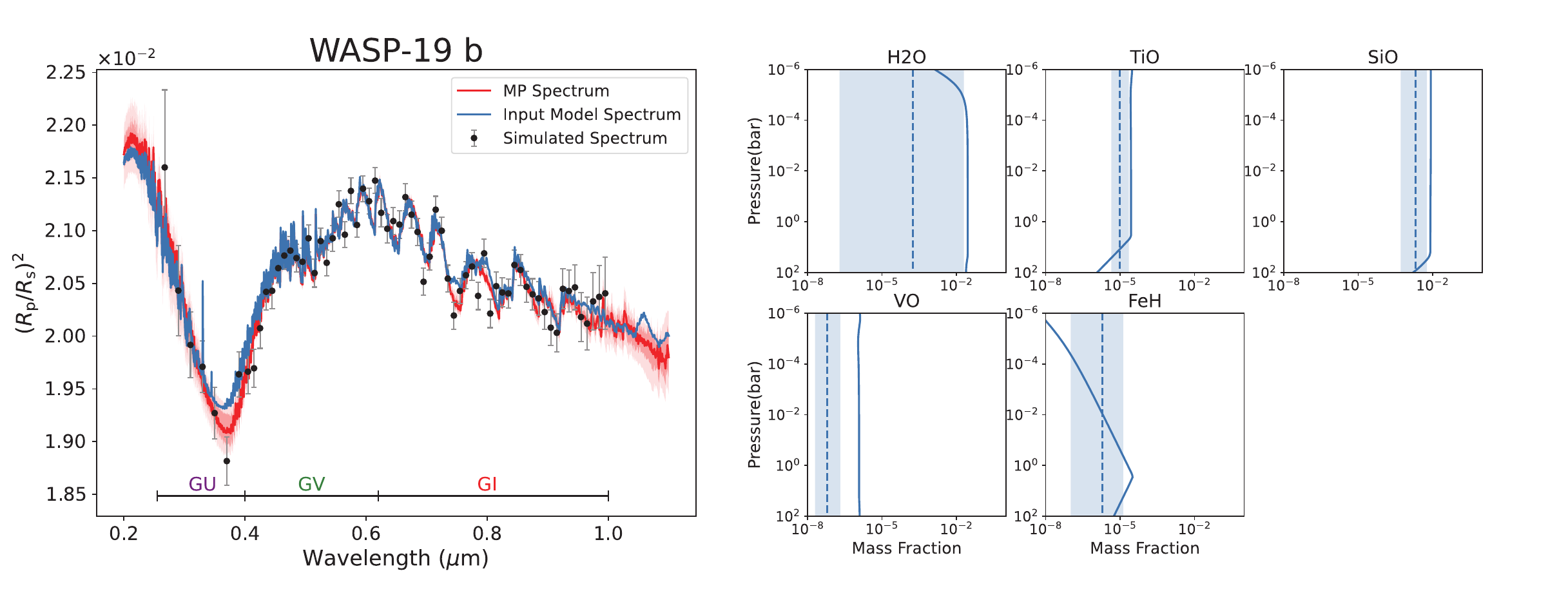}
\caption{Free-chemistry retrieval results for WASP-19\,b. Left panel: The transmission spectrum derived from simulated CSST transit observation consisting of two partial transits taken in two visits per band (black filled circles with error bars), together with the MP model (red line) and the input theoretical planet spectrum (blue line). The shaded regions represent the 68\% and 95\% credible intervals of the posterior spectra. Right panel: Retrieved vertical profiles of the species in mass fractions, compared with the calculated EQ-chemistry abundances. The blue dashed curves represent the posterior medians, and the shaded regions denote the 16th–84th percentile intervals, while the blue solid lines represent the mass fraction profiles from the input models.}
\label{fig:free-chem-wasp19}
\end{figure*}

\begin{figure*}
\centering
\includegraphics[width=1\textwidth]{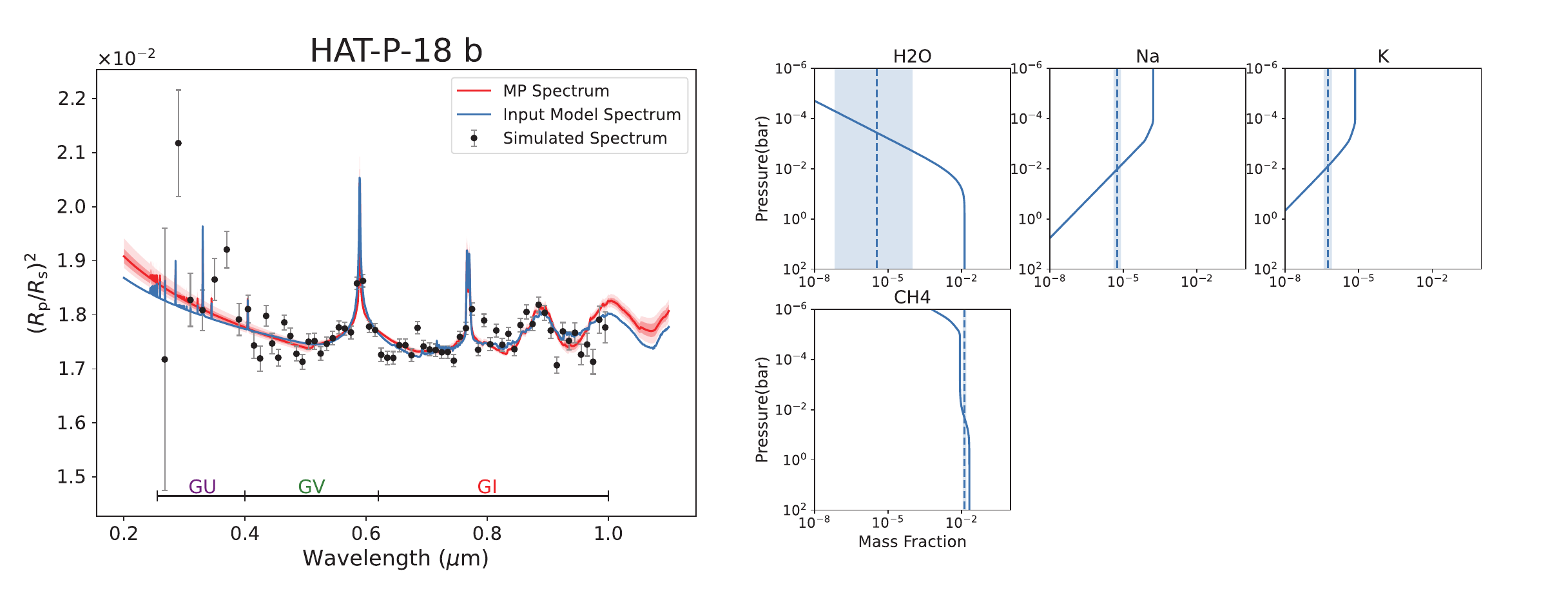}
\caption{Same as Fig.\ref{fig:free-chem-wasp19}, but for HAT-P-18\,b.}
\label{fig:free-chem-hat-p-18}
\end{figure*}

\begin{figure*}
\centering
\includegraphics[width=1\textwidth]{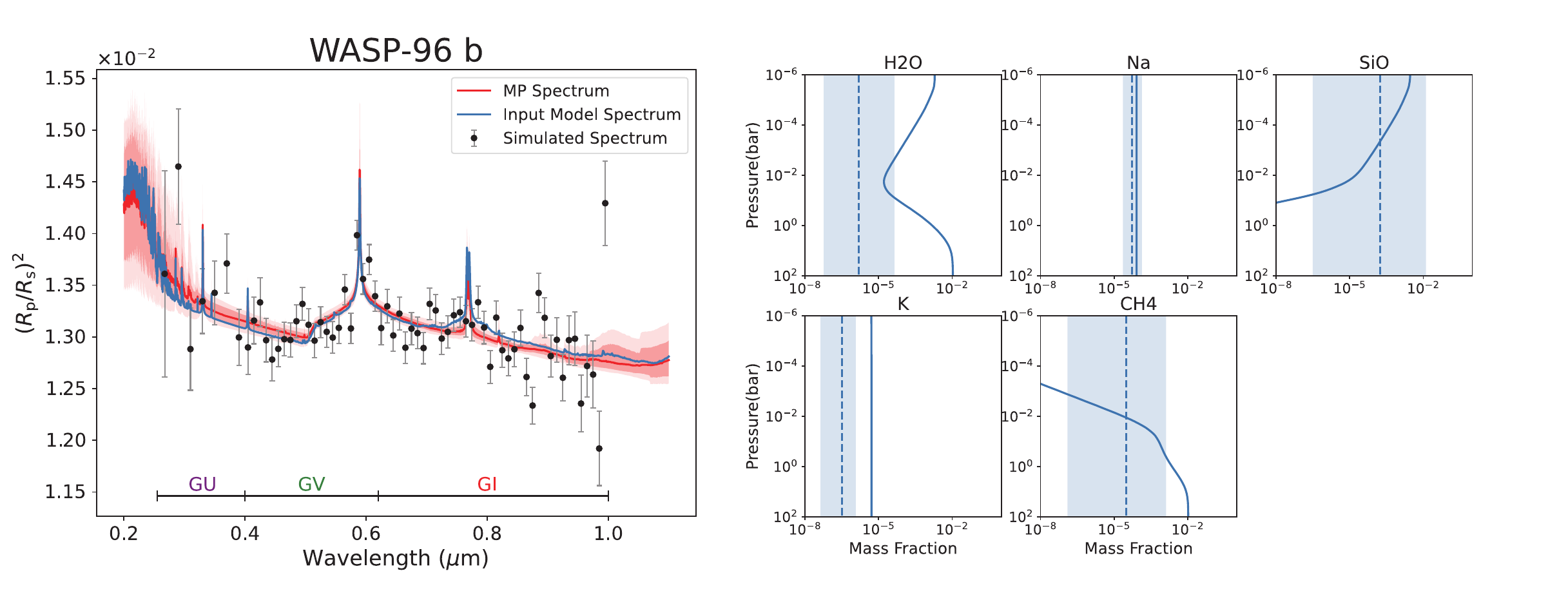}
\caption{Same as Fig.\ref{fig:free-chem-wasp19}, but for WASP-96\,b.}
\label{fig:free-chem-wasp96}
\end{figure*}

\begin{figure*}
\centering
\includegraphics[width=1\textwidth]{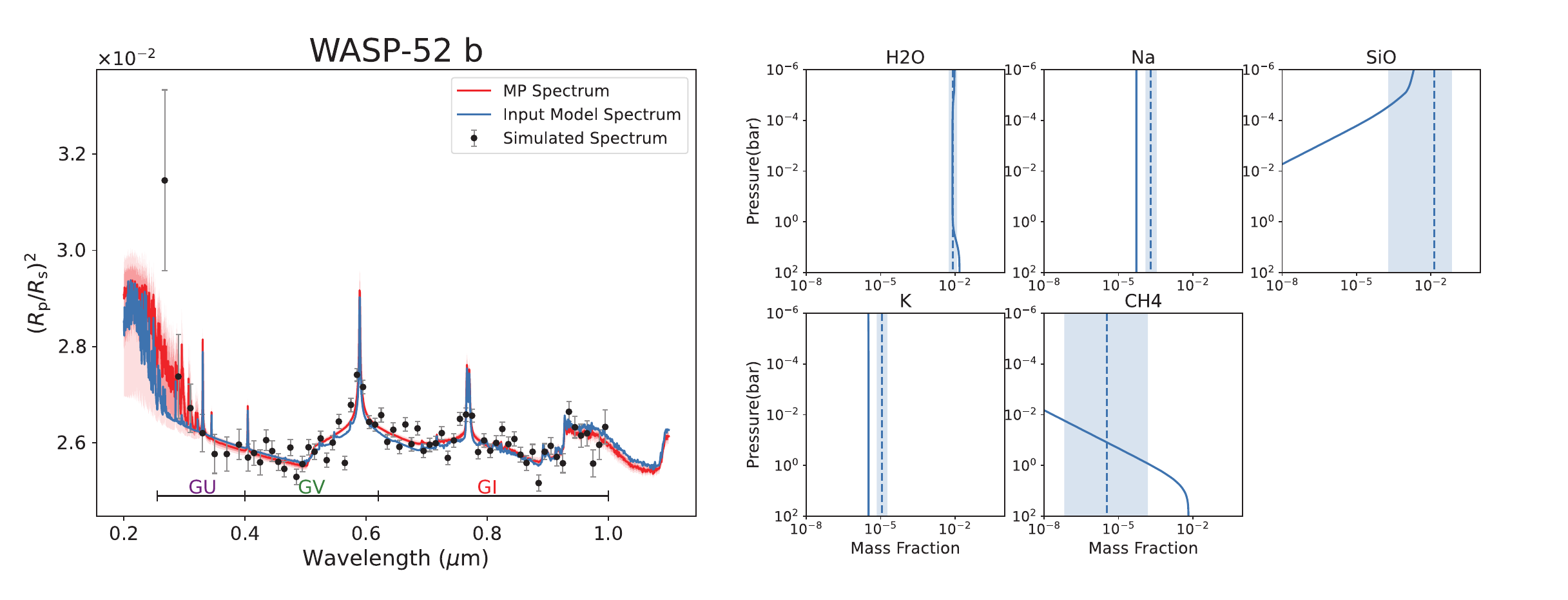}
\caption{Same as Fig.\ref{fig:free-chem-wasp19}, but for WASP-52\,b.}
\label{fig:free-chem-wasp52}
\end{figure*}

\begin{figure*}
\centering
\includegraphics[width=1\textwidth]{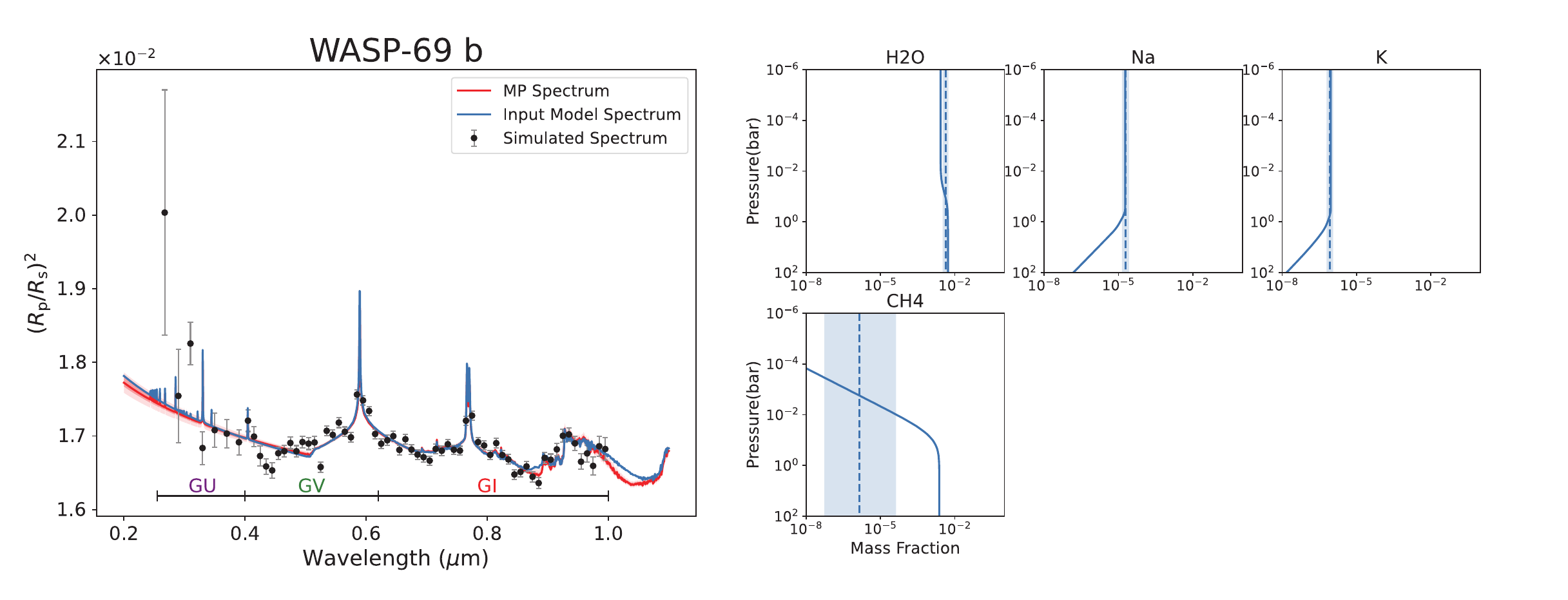}
\caption{Same as Fig.\ref{fig:free-chem-wasp19}, but for WASP-69\,b.}
\label{fig:free-chem-wasp69}
\end{figure*}

\begin{figure*}
\centering
\includegraphics[width=1\textwidth]{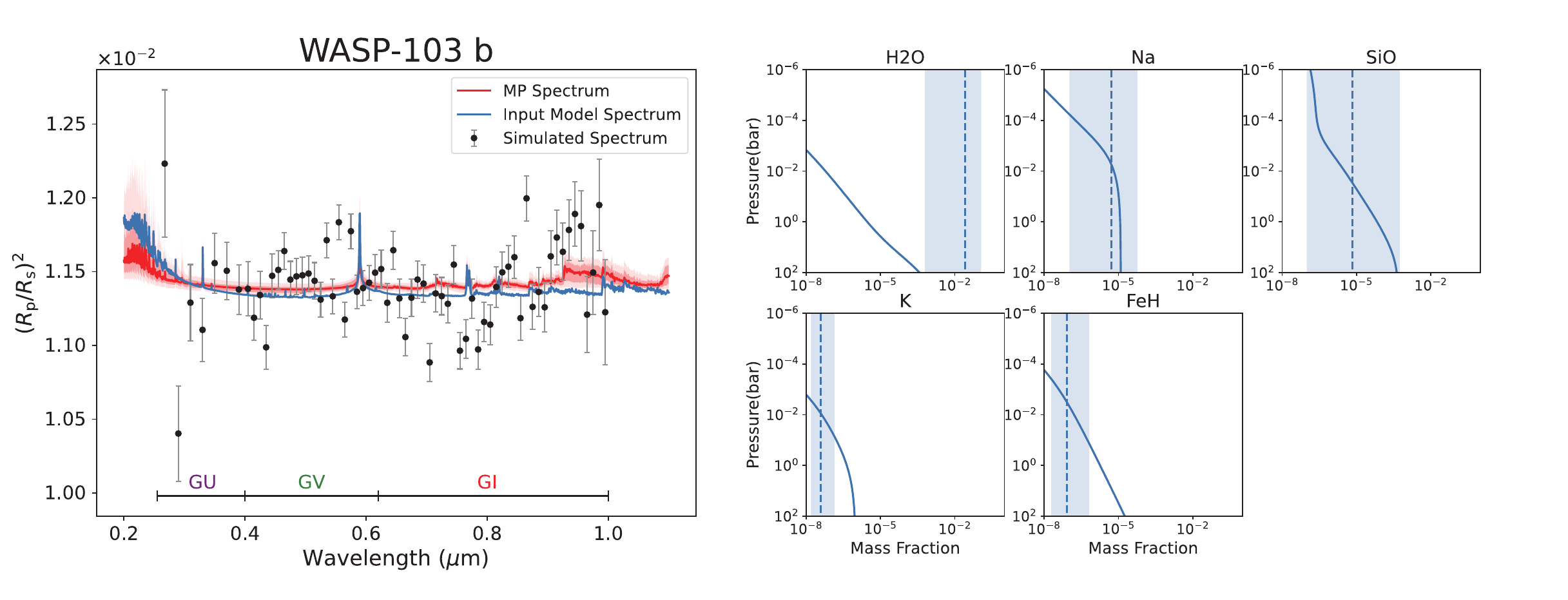}
\caption{Same as Fig.\ref{fig:free-chem-wasp19}, but for WASP-103\,b.}
\label{fig:free-chem-wasp103}
\end{figure*}

Since the theoretical spectra are generated under the assumption of chemical equilibrium, the EQ retrievals in Section \ref{sec:chem-eq} naturally converge toward solutions close to the input models. In real observations, however, EQ chemistry does not always provide satisfactory fits, and free-chemistry retrievals — in which the abundances of individual species are independently retrieved — also need to be considered, and are performed for all the 6 planets in this study. The retrieved results are shown in Figures~\ref{fig:free-chem-wasp19}–\ref{fig:free-chem-wasp103} for each planet, and summarized in Table~\ref{tab:chem-free}. 

For each of Figures~\ref{fig:free-chem-wasp19}–\ref{fig:free-chem-wasp103}, the left panel compares for each planet the model spectrum generated using the MPs (the red line) and the deduced transmission spectrum from simulated CSST observation (the black-filled circles) against the input theoretical transmission spectrum (the blue line), while the right panel gives the retrieved abundance profiles from the free-chemistry retrievals. The retrieved planetary radii, temperatures, and mass fractions of selected chemical species are summarized in Table~\ref{tab:chem-free}. Only species that produce significant spectral signatures within the CSST wavelength range are included in the retrieval models. This selection is guided by the expected detectability of each species given the wavelength coverage, spectral resolution, and noise level of the simulated observations. Including additional species that do not produce measurable features would increase the number of free parameters without improving the fit, and may introduce degeneracies in the retrieval. Therefore, species with negligible spectral impact are not included.

\begin{table*}[htbp]
 %\textbf{Free-chemistry retrieval results}
  \caption{Retrieved atmospheric parameters for the free-chemistry cases. The quoted uncertainties correspond to the 16th and 84th percentiles of the posterior distributions. 
  Abundances are given as the log mass fractions. Species marked with “–” lacked significant spectral signatures and were therefore not included in retrievals.}
  \label{tab:chem-free}
    \centering
    %\resizebox{\textwidth}{!}{
    \tiny
    \begin{tabular}{ccccccccccc}
    \hline
    \noalign{\smallskip}
    Planet Name & $R$($R_{\mathrm{jup}}$) & $T$(K) & Na & K & H$_2$O & CH$_4$ & TiO & VO & SiO & FeH \\
    \noalign{\smallskip}
    \hline
    \noalign{\smallskip}
    WASP-19\,b & 1.385 $^{+0.007}_{-0.008}$ & 2767 $^{+146}_{-178}$ & - & - & -3.76 $^{+2.06}_{-2.95}$ & - & -5.02 $^{+0.37}_{-0.34}$ & -7.20 $^{+0.52}_{-0.50}$ & -2.70 $^{+0.47}_{-0.59}$ & -5.71 $^{+0.84}_{-1.28}$ \\
    HAT-P-18\,b & 0.998 $^{+0.002}_{-0.002}$ & 1123 $^{+86}_{-85}$ & -5.25 $^{+0.15}_{-0.15}$ & -6.25 $^{+0.16}_{-0.17}$ & -5.46 $^{+1.47}_{-1.71}$ & -1.88 $^{+0.07}_{-0.07}$ & - & - & - & -\\
    WASP-96\,b & 1.203 $^{+0.004}_{-0.004}$ & 1355 $^{+254}_{-276}$ & -4.28 $^{+0.41}_{-0.36}$ & -6.50 $^{+0.57}_{-0.87}$ & -5.80 $^{+1.45}_{-1.44}$ & -4.52 $^{+1.63}_{-2.39}$ & - & - & -3.75 $^{+1.86}_{-2.75}$ & -\\
    WASP-52\,b & 1.268 $^{+0.002}_{-0.002}$ & 1261 $^{+99}_{-103}$ & -3.70 $^{+0.24}_{-0.21}$ & -4.94 $^{+0.21}_{-0.21}$ & -2.09 $^{+0.18}_{-0.16}$ & -5.47 $^{+1.65}_{-1.72}$ & - & - & -1.86 $^{+0.71}_{-1.86}$ & -\\
    WASP-69\,b & 1.056 $^{+0.001}_{-0.001}$ & 846 $^{+56}_{-54}$ & -4.72 $^{+0.15}_{-0.13}$ & -6.07 $^{+0.13}_{-0.14}$ & -2.37 $^{+0.12}_{-0.12}$ & -5.84 $^{+1.47}_{-1.43}$ & - & - & - & - \\
    WASP-103\,b & 1.526 $^{+0.002}_{-0.002}$ & 990 $^{+415}_{-263}$ & -5.29 $^{+1.07}_{-1.69}$ & -7.42 $^{+0.56}_{-0.39}$ & -1.60 $^{+0.66}_{-1.61}$ & - & - & - & -5.17 $^{+1.92}_{-1.83}$ & -7.09 $^{+0.91}_{-0.63}$\\
    \noalign{\smallskip}
    \hline
    %曝光时间
    \end{tabular}
    %}

\end{table*}

Compared with the EQ results, the free-chemistry retrievals generally provide fits to the simulated spectra that are equal to or slightly improved, owing to the increased number of free parameters, which allow larger flexibility in reproducing spectral features. However, this additional flexibility also leads to broader posterior distributions for individual parameters, particularly for temperature and molecular abundances.

For planets with relatively high TSM values, i.e., WASP-19\,b, WASP-69\,b, WASP-52\,b, and HAT-P-18\,b, the free-chemistry retrievals recover the dominant absorbers with still reasonable accuracy. In particular, the abundances of strong optical absorbers such as Na and K are well constrained for most targets, while metal oxides (e.g., TiO/VO) are well recovered only for WASP-19\,b in which the spectral signatures are prominent. Molecular species that contribute less to the observed spectra possess larger uncertainties but remain broadly consistent with their input abundance levels.

In contrast, for lower-TSM planets such as WASP-103\,b and WASP-96\,b, the free-chemistry retrievals show significantly increased uncertainties in molecular mass fractions. While the mass fraction of the alkali species in WASP-96\,b remains well constrained, most other species exhibit significantly broader posterior distributions, and all the considered molecules are just marginally detected with about $3\sigma$ detection significance, or barely detected with less than $3\sigma$ significance. For WASP-103\,b, the spectrum is largely featureless due to strong continuum opacity, resulting in weak or non-detections of molecular species.

Overall, relaxing the assumption of chemical equilibrium to free chemistry increases uncertainties of radius and temperature, particularly for planets with modest TSM values. Nevertheless, strong absorbers remain robustly detectable within the CSST wavelength coverage.

\section{Discussion}
\label{sec:discussion}

\subsection{\texorpdfstring{WASP-19\,b}{WASP-19 b}}
\label{subsec:W19b}

Overall, the simulated transmission spectrum (the black filled circles) of WASP-19\,b follows the general trend predicted by the EQ theoretical spectrum, as shown in Panel a of Fig.~\ref{fig:chem-eq-retrieval} and Fig.~\ref{fig:free-chem-wasp19}, but with a significantly deeper absorption near $\sim$0.4$\mu$m, which may be caused by smaller SNR at this spectroscopic bin. In the UV, the spectrum is mostly shaped by a combination of SiO absorption features and continuum opacity, although both equilibrium- and free-chemistry retrievals tend to overestimate the UV spectral slope slightly and overestimate the absorption depth near $\sim0.4\mu$m. 

As a result, the retrieved planet temperature is overestimated due to stronger molecular absorption, leading to an underestimation of the continuum level (and consequently of the planetary radius), to compensate for the input transmission spectral level. The bulk C/O ratio is weakly constrained, due to the absence of strong carbon-bearing molecular features within the CSST wavelength coverage. Similarly, because molecular absorption increases with temperature, free-chemistry retrievals yield systematically lower molecular abundances due to an overestimated retrieved temperature. These discrepancies highlight that an accurate determination of the continuum opacity is critical for robust constraints on the atmospheric structure and overall chemical composition.

Previous HST transit observation of WASP-19\,b revealed a largely featureless transmission spectrum in the optical and clear H$_2$O absorption in the NIR~\citep{Huitson2013}. Subsequent ground-based optical observations reported tentative evidence of TiO features, together with enhanced short-wavelength scattering~\citep {Sedaghati2017}. Later, optical transmission spectroscopy with Magellan/IMACS again reported a featureless spectrum~\citep {Espinoza2019}. More recently, high-resolution ESPRESSO spectroscopy detected only a marginal TiO signal consistent with very low abundances, while independently confirming strong optical scattering~\citep{Sedaghati2021}. 

All these discoveries, the lack of absorption in the optical, weak signals in the NIR, and strong blue scattering, are consistent with our predicted transmission spectrum, as shown in Fig.~\ref{fig:chem-eq-retrieval} and Fig.~\ref{fig:free-chem-wasp19}, confirming our model setup is appropriate. Therefore, the predicted strong SiO absorption is likely a robust feature of the model, although it awaits confirmation from future UV observations, which will become accessible with CSST in the near future, and be well characterized by Tianlin~\citep{Wang2023a,Wang2023b} in the 2040s. 

\subsection{\texorpdfstring{HAT-P-18\,b}{HAT-P-18 b}}

HAT-P-18\,b is a warm Saturn-mass planet orbiting a K-type star. Our simulated transmission spectrum agrees closely with the EQ  model, and the EQ retrieval accurately recovers the atmospheric parameters. The largest deviation occurs at the shortest GU wavelengths in the free-chemistry case, where the relatively low photon flux from the K-dwarf host leads to larger uncertainties and a slight offset from the theoretical curve. This causes the free-chemistry retrieval to infer a steeper continuum slope and, consequently, a mildly overestimated temperature. In the GI band, the overlap of H$_2$O and CH$_4$ features limits their reliable detection and abundance measuring at low resolution, though CH$_4$ remains reasonably well constrained. The retrieved Na and K abundances are consistent with the calculated EQ-chemistry values at the reference pressure level ($10^{-2}$ bar). Overall, the simulations indicate that future CSST observations could effectively constrain the atmospheric structure and chemical abundances of HAT-P-18\,b.

Previous optical transmission spectroscopy revealed a strong blueward scattering slope and no significant \ion{Na}{1} absorption, suggesting haze-dominated opacity~\citep{Kirk2017}, while JWST observations confirmed molecular absorbers such as H$_2$O and CO$_2$ in the near-infrared~\citep{Fournier2024}. In contrast, our cloud-free EQ simulations predict clearly detectable Na and K lines, as well as measurable CH$_4$ features within the CSST wavelength range. Their absence would indicate a weakly featured transmission spectrum, consistent with an effectively opaque atmosphere (e.g., due to clouds or hazes). Future UV-to-NIR, higher-sensitivity observations are therefore crucial for characterizing the aerosols involved and confirming that alkali and methane features are suppressed by haze opacity in this low-gravity atmosphere.

\subsection{\texorpdfstring{WASP-96\,b}{WASP-96 b}}

WASP-96\,b, the one with the second lowest TSM in the studied sample, exhibits noticeable discrepancies between our simulation and the input model for both the EQ- and free-chemistry retrievals, which are within the uncertainties of our simulated spectrum. In general, both retrievals can reproduce the absorption features and continuum in the GV band well, but less so in the GU and GI bands. Biases at the continuum level lead to a slight overestimation of the $r_{\rm p}$ and $T$, which in turn enhance molecular absorption and result in an underestimation of [Fe/H]. Again, the C/O ratio remains weakly constrained owing to the limited sensitivity to carbon-bearing species. The free-chemistry retrieval more closely follows the overall spectral shape: the Na abundance is accurately recovered, whereas K is mildly underestimated due to deviations in the line profile. Other molecular abundances are poorly constrained but consistent with the input values within uncertainties. Note that the sharp rise near 1.0$\mu$m may be due to the reddest simulated data point, which is significantly larger than adjacent data points. In future CSST observations, we may consider excluding such data points, as they are outliers with large uncertainties.

Previous transmission spectroscopy studies indicate that WASP-96\,b hosts a relatively clear atmosphere showing prominent alkali absorption, consistent with our prediction. Ground-based VLT observations detected a pressure-broadened Na feature~\citep{Nikolov2018}, later confirmed by joint optical and HST near-infrared data that also revealed the presence of H$_2$O~\citep{McGruder2022}. JWST Early Release Observations further provided precise constraints on H$_2$O, CO$_2$, and K, while suggesting weak short-wavelength aerosol scattering~\citep{Taylor2023}. Overall, existing observations support a clear or low-cloud atmosphere with strong alkali absorptions. CSST UV-VIS measurements would provide independent constraints on alkali line cores, particularly the optical scattering slope, thereby refining abundance estimates and probing residual aerosol opacity.

\subsection{\texorpdfstring{WASP-52\,b}{WASP-52 b}}

The simulations of WASP-52\,b yield synthetic spectra that are consistent with the input theoretical model, particularly in the GV and GI bands. However, clear deviations occur in the GU band, where the derived transit depths at the shortest wavelengths exceed theoretical predictions. This may be caused by the late spectral type of the host star, similar to the case of HAT-P-18\,b. EQ retrievals yield quite accurate constraints on $R_p$ and $T$ and a slight overestimate of [Fe/H] due to the above-mentioned upwards bias in the GU band that may amplify SiO absorption signals. We note that our input equilibrium-chemistry model neglects rainout and condensation, which may overestimate the abundance of SiO in the upper atmosphere. Again, the lack of carbon-bearing molecular features within CSST's wavelength coverage leads to poorly constrained C/O ratios.  In free chemistry retrievals, the mass fractions of Na, K, and H$_2$O are well constrained, albeit with slightly higher values for Na and K, because the simulated spectral line wings are broader than the input ones. While retrieved abundances of SiO and CH$_4$ are consistent with the theoretical input, but with a wide posterior distribution. 

WASP-52\,b is a highly-inflated HJ, with previous high-resolution observations from VLT/ESPRESSO revealing significant Na, K, and H$\alpha$ absorption~\citep{Chen2020}. However, quantitative abundance retrievals are largely hampered by the intensive magnetic activity and star-spot crossings of its K-type host star~\citep{Bruno2020}. Our simulation results demonstrate that future CSST slitless spectroscopy could effectively characterize key alkali-metal and water features in the GV and GI bands.

\subsection{\texorpdfstring{WASP-69\,b}{WASP-69 b}}

The simulated transmission spectra of WASP-69\,b for the EQ- and free-chemistry cases agree well with theoretical models, probably benefiting from its high TSM. The EQ retrievals accurately recover all physical parameters, with the MP spectrum closely matching the input spectrum. Similarly, the free-chemistry retrievals provide robust constraints on the mass fractions of H$_2$O, Na, and K. The constraint on CH$_4$ is marginal, which is reasonable as it does not have strong features in the CSST bands. 

WASP-69\,b is found to have an extensively escaping atmosphere, notably evidenced by a long, comet-like tail of metastable helium~\citep{Nortmann2018}. Previous studies have reported the detection of Na~\citep{Casasayas2017} and the tentative detection of TiO~\citep {Ouyang2023} using a ground-based telescope, while a scattering haze may have altered the optical slope, complicating abundance retrievals~\citep{Murgas2020}. Our simulations show that future CSST slitless spectroscopy could, in principle, provide high-confidence measurements of alkali lines and the optical continuum, which are essential for constraining haze properties. In addition, WASP-69\,b represents an excellent target for repeated broadband monitoring to study atmospheric variability and mass-loss processes in details.

\subsection{\texorpdfstring{WASP-103\,b}{WASP-103 b}}

WASP-103\,b shows relatively large dispersion in the simulated transmission spectrum, because of  its lowest TSM and [Fe/H] in our sample, and thus intrinsically weak spectral features and large error bars. In addition, the inclusion of strong continuum opacity from H$^-$ further suppresses molecular and atomic absorption features, resulting in an overall nearly featureless transmission spectrum within the CSST wavelength range. As a result, the EQ retrieval deviates noticeably from the theoretical model with large uncertainties, particularly in $T$ and [Fe/H]. The free-chemistry retrieval does not significantly improve the fit, as the lack of distinct spectral features limits its ability to constrain molecular abundances, leading primarily to upper limits rather than robust detections.

Previous studies on WASP-103\,b reported discrepant optical transmission results: ground-based observations suggested possible Na and K absorption~\citep{Lendl2017}, whereas later optical and HST near-infrared data indicated a largely featureless spectrum~\citep{Wilson2020}.
Our simulations suggest that, under the influence of strong continuum opacity, future CSST UV–VIS observations may also yield a largely featureless spectrum. Increasing the number of observed transits (see Section~\ref{sec:discussion-4transit}) could improve the signal-to-noise ratio and help place tighter constraints on possible Na and K absorption.

\subsection{Impact of multiple transit observations and shorter orbit duration}
\label{sec:discussion-4transit}

\begin{figure*}
    \centering
    \includegraphics[width=1\textwidth]{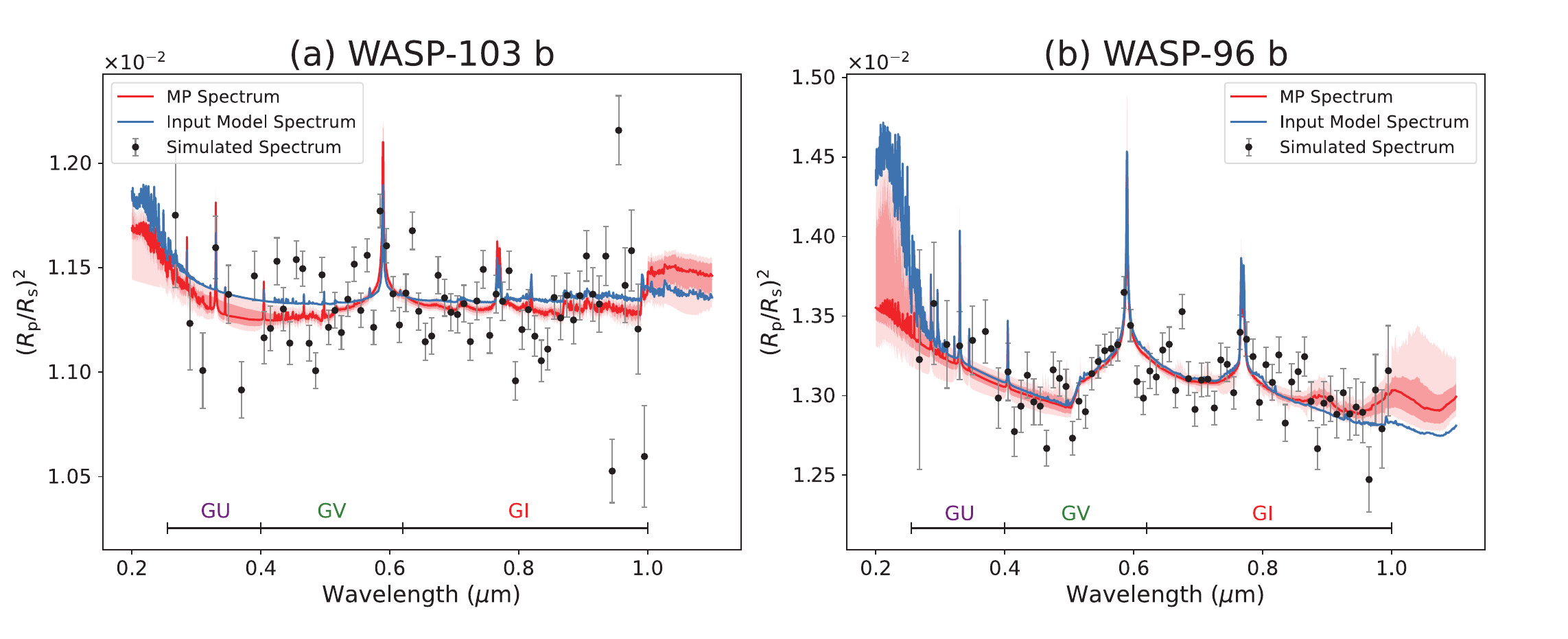}
    \caption{EQ Retrieval results of WASP-103\,b and WASP-96\,b with 4 visits per band.}
    \label{fig:4-transit}
\end{figure*}

To obtain better spectral retrievals and a more reliable determination of planet atmospheric parameters, we have conducted similar simulations covering multiple transits to achieve higher SNRs. The simulations are run for WASP-103\,b and WASP-96\,b, each incorporating two additional transits (in six orbits) beyond the baseline scenario. As illustrated in Fig.\ref{fig:4-transit}, the addition of new observations significantly reduces spectral scatter and tightens the posterior distributions. In both cases, the retrieved equilibrium parameters approach the theoretical input values more closely, with smaller uncertainties in planetary radius, temperature, and metallicity. 
The improvement is especially notable for WASP-103\,b, where the initially weak spectral features led to large dispersion in single-transit simulations. With increased SNR from multiple transits, weak spectral features begin to emerge from the noise, partially reducing the degeneracies seen in the single-transit case. In particular, the Na absorption feature becomes marginally detectable in the four-transit scenario, although the overall spectrum remains largely featureless due to strong continuum opacity. These results demonstrate that multiple-transit observations can effectively mitigate noise-induced biases in the continuum level and atmospheric parameters, enabling more robust abundance constraints, particularly for planets with low TSM or intrinsically weak spectral signatures.

%\ww{--------------------modification stopped here in 20260329---------------} %%%%%%%%%%

Another point is that the available continuous observing time per orbit may be shorter than 40 minutes, limited by CSST's low orbit and shutdown due to safety regulations. This means that to cover one full planet transit and baselines, more CSST orbits are required. This segmented coverage can introduce additional systematics due to orbit-to-orbit variations in instrument conditions (e.g., thermal environment and pointing), which complicate the combination of data from different orbits. To quantify the sensitivity of our results to this effect, we performed two additional simulations assuming reduced continuous observing windows of 30 minutes and 20 minutes per orbit, respectively. These tests were carried out for two representative targets, WASP-19\,b and WASP-52\,b, which span different temperatures and TSMs. The simulated spectra were generated and analyzed following the same procedures described above. As shown in Appendix~\ref{sec:app-a}, all the retrievals are successful in deriving planetary atmospheric parameters reliably, including C/O, [Fe/H] for the EQ case, and mix ratios of the individual species for the free-chemistry case.

Furthermore, to assess the robustness of the simulated retrieval results to random noise realizations, we repeated the CSST observation simulations multiple times for WASP-19\,b with different random seeds. This procedure allows us to assess the dispersion in the retrieved atmospheric parameters arising purely from stochastic noise, while keeping all instrumental and observational configurations fixed. We find that the random noise does not induce any bias and the simulation is robust, as summarized in Section~\ref{sec:random}.

\subsection{\texorpdfstring{WASP-178\,b}{WASP-178 b}: A test case for time-correlated noise and comparison between CSST and  HST}\label{sec:W178}

%\subsection{WASP-178\,b}

All the simulations above consider only white noise and do not include time-correlated (red) noise when generating observed CSST spectra and deriving the transmission spectra. That may be one reason that CSST will potentially achieve substantially smaller scatter and uncertainties than HST. To obtain a conservative assessment of the ''real'' power of CSST as compared to HST, we injected red noise at a level exceeding that of the HST data for WASP-178\,b, a planet previously observed by HST, to assess the impact of time-correlated noise on CSST transmission spectroscopy. 

To construct a realistic input transmission spectrum for WASP-178,b, we first perform a retrieval on the published HST data following \citet{Lothringer2022} (hereafter L22). The retrieval adopts a hybrid scheme in which selected species are treated under free chemistry, while the remaining species follow equilibrium chemistry. The retrieved abundances are allowed to vary over a uniform prior range of -12 - 0 dex. The opacity contribution from H$^-$ is also included due to the high $T_{\mathrm{eq}}$ We assume an isothermal temperature profile, which maintains consistency with other atmosphere models in this work and differs from the temperature structure adopted in L22.

The resulting best-fit spectrum shows a prominent SiO absorption feature in the UV, while the spectrum at longer wavelengths is largely featureless and close to a flat continuum. In the subsequent CSST simulations, we assumed only a single transit observation (5 orbits, same to HST) for each band and adopt exactly the same retrieval setup as used for the HST-based reconstruction to ensure consistency, even though the spectral features outside the UV remain weak.

\begin{figure*}[htbp]
    \centering
    \includegraphics[width=1\textwidth]{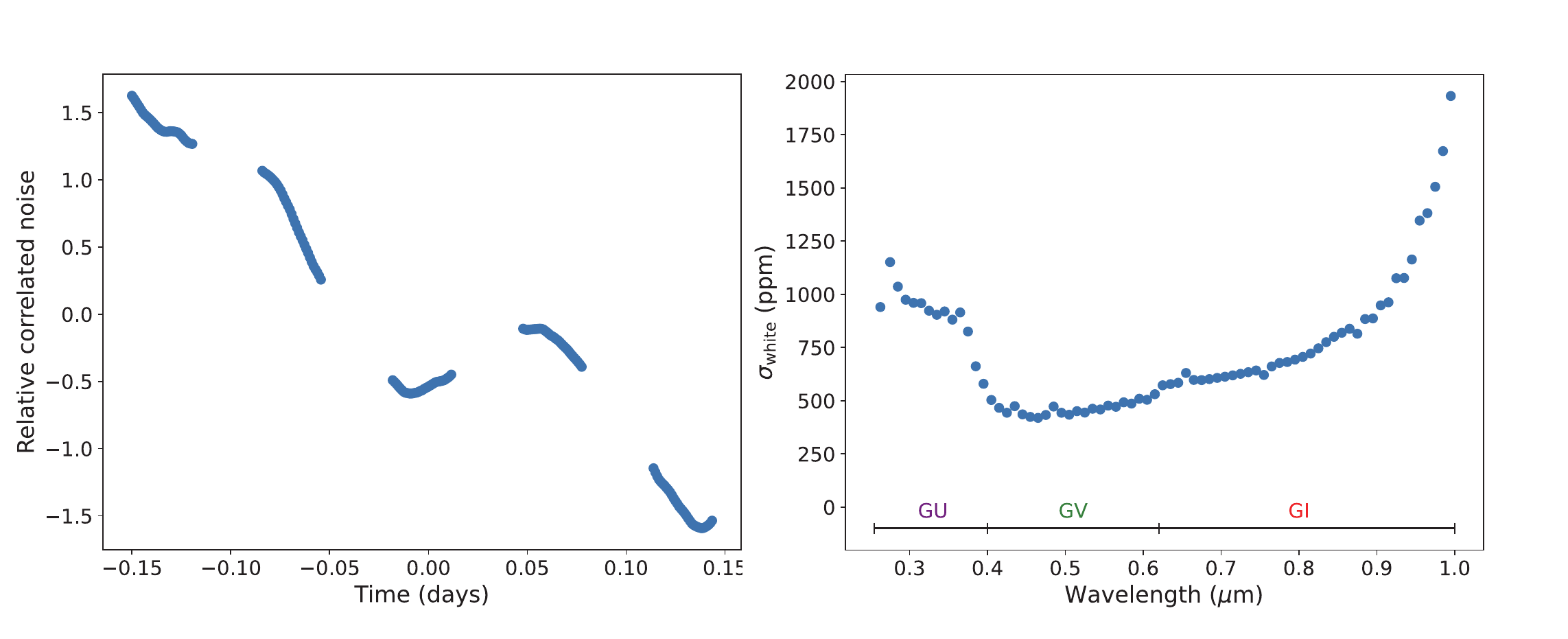}
    \caption{Left: Amplitude of injected red noise generated by GP, normalized to zero mean and unit variance. Right: Median white noise level in each wavelength bin.}
    \label{fig:yerr}
\end{figure*}

The time-correlated noise is modeled using the Gaussian process (GP) package \texttt{george}\citep{Ambikasaran2015} with a Matérn-3/2 kernel. We adopt a characteristic timescale of 30 minutes, which is comparable to the expected timescales of instrumental systematics in low-Earth-orbit observations. The trend of injected noise is shown in the left panel of Fig.~\ref{fig:yerr}, with the amplitude ($\sigma_{\mathrm{red}}$) scaled with the median white-noise level ($\sigma_{\mathrm{white}}$ in right panel of Fig.~\ref{fig:yerr}) in each wavelength bin. As described later, after Gaussian-process detrending, most of the correlated noise is removed.

We note that this treatment represents a simplified description of time-correlated noise. In practice, additional systematics may arise from instrument-specific effects or stellar variability, which can introduce wavelength-dependence. Such effects are not explicitly included in the present simulations. A more comprehensive treatment incorporating these sources of systematics is beyond the scope of this work and will be explored in future studies.

\begin{figure*}[htbp]
\centering
\gridline{\fig{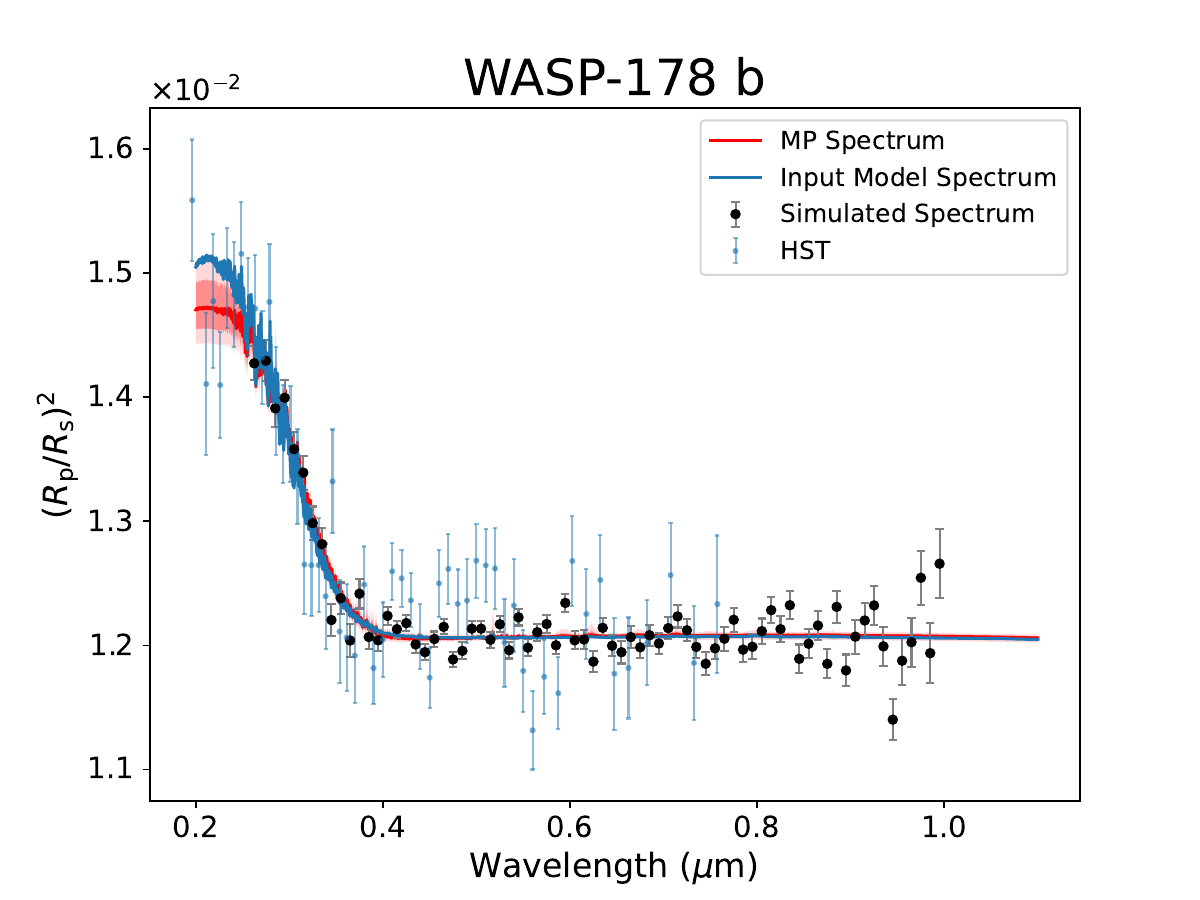}{0.5\textwidth}{(a) Without red noise}
\hspace{-0.05\textwidth}
\fig{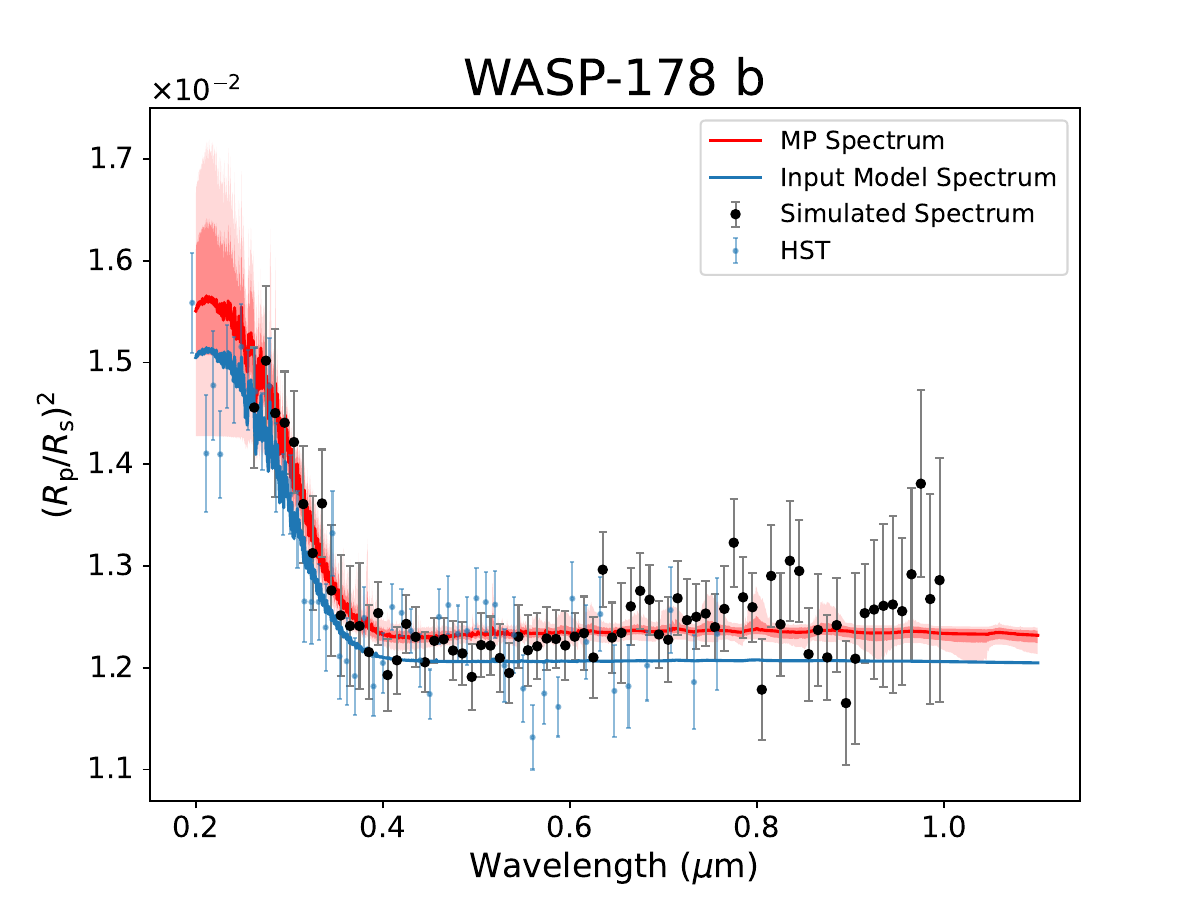}{0.5\textwidth}{(b) $\sigma_{\mathrm{red}} = \sigma_{\mathrm{white}}$}
}
\gridline{\fig{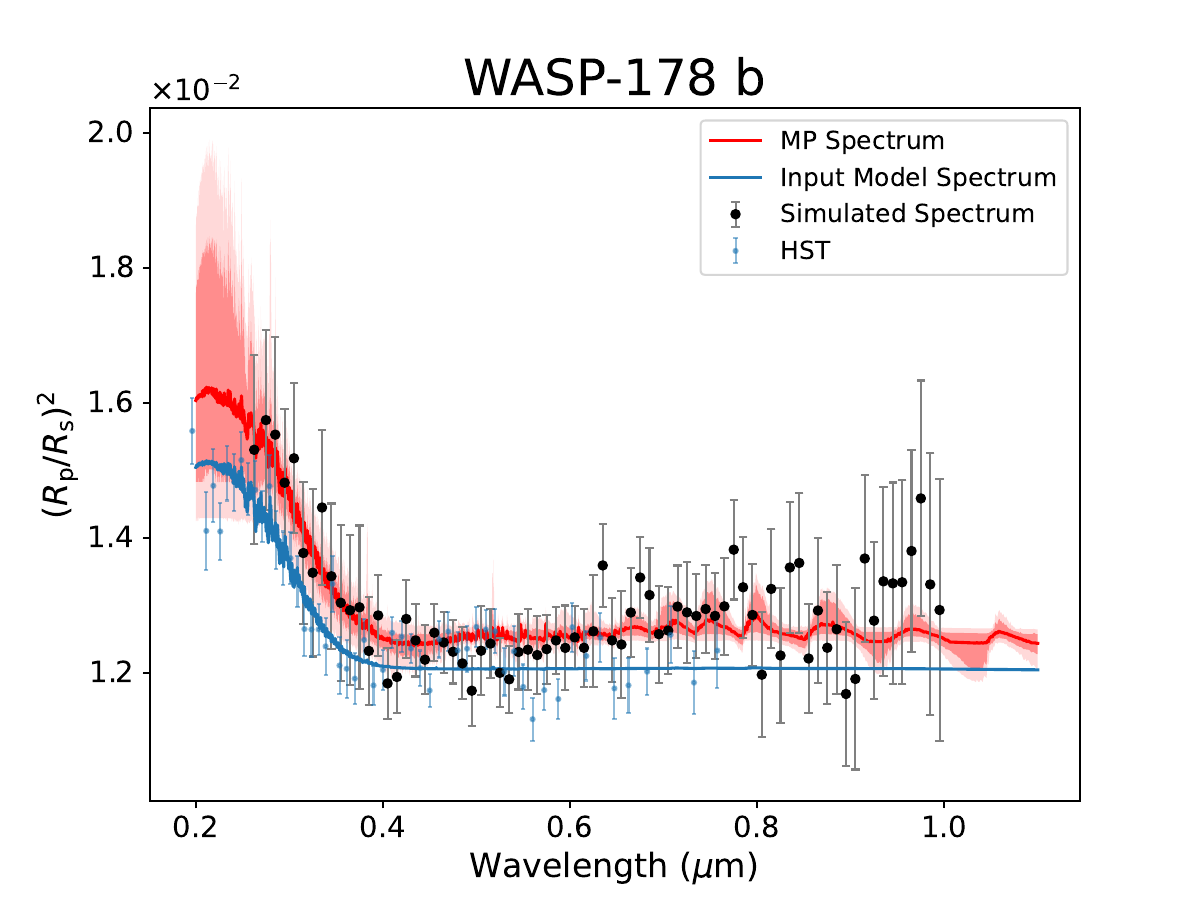}{0.5\textwidth}{(c) $\sigma_{\mathrm{red}} = 3 \times \sigma_{\mathrm{white}}$}
\hspace{-0.05\textwidth}
\fig{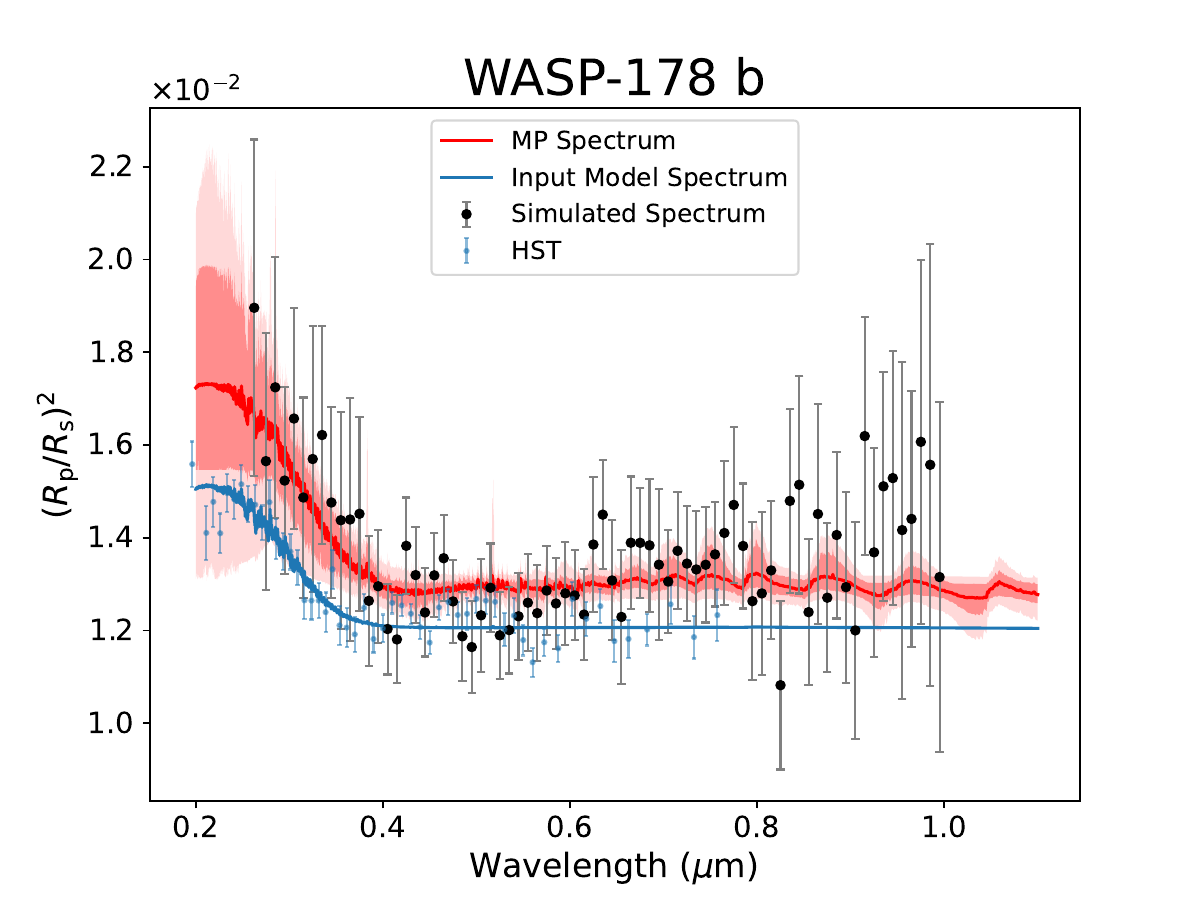}{0.5\textwidth}{(d) $\sigma_{\mathrm{red}} = 10 \times \sigma_{\mathrm{white}}$}
}
\caption{Comparison between simulated CSST transmission spectra (black dots) and retrieved MP spectrum (red lines) of WASP-178\,b without red noise (a) and with injected different levels red noise modeled using Gaussian processes (b)(c)(d) and HST observations (blue  dots). The blue line shows the theoretical spectra derived from HST observations and used as input to CSST simulations. These simulations cover a single transit for each band.}
\label{fig:wasp-178b-compare}
\end{figure*}

Following the same procedure, we first analyzed simulated light curves containing only white noise. As shown in Fig.~\ref{fig:wasp-178b-compare}(a), in the absence of red noise, the simulated CSST observations exhibit substantially smaller scatter and uncertainties than the corresponding HST data, reflecting the expected photon-noise performance of CSST. CSST will use optical/UV CCD detectors and fixed filter/grating elements for its slitless spectroscopic bands, so its detector-related systematics may differ from the transient effects commonly seen in HST/WFC3 near-infrared time-series observations. However, orbit-related systematics may still be present and will need to be characterized after launch.

After injecting time-correlated (red) noise, the spectroscopic light curves display evident long-term trends. We therefore incorporated GP into the MCMC fitting of each spectroscopic channel. The reconstructed transmission spectrum is shown in Fig.~\ref{fig:wasp-178b-compare}(b). Even with relatively strong red noise, the resulting spectral uncertainties remain comparable to those of HST. The overall spectral morphology is preserved, including the pronounced upward slope in the GU band and featureless in the GI region, both of which are consistent with the theoretical spectrum.

\begin{table*}[htbp]
    \textbf{Atmospheric retrieval results for WASP-178\,b}
    \centering
    \tiny
    \begin{tabular}{ccccccccccc}
    \noalign{\smallskip}
    \hline
    \noalign{\smallskip}
    Retrieval Name & $R(R_{\mathrm{jup}})$ & $T$(K) & SiO & TiO & VO & Fe & Fe II & Mg & Mg II & [Fe/H]\\
    \noalign{\smallskip}
    \hline
    \noalign{\smallskip}
    HST & 1.838$^{+0.012}_{-0.015}$ & 2778$^{+224}_{-145}$ & -2.20$^{+0.32}_{-0.48}$ & -8.17$^{+2.67}_{-2.36}$ & -5.45$^{+4.07}_{-4.35}$ & -6.47$^{+3.43}_{-3.34}$ & -3.99$^{+2.42}_{-4.92}$ & -10.31$^{+1.25}_{-1.09}$ & -9.75$^{+1.52}_{-1.49}$ & -0.82$^{+1.20}_{-1.36}$\\
    No red noise & 1.823$^{+0.010}_{-0.011}$ & 2729$^{+100}_{-82}$ & -1.62$^{+0.24}_{-0.28}$ & -7.66$^{+2.47}_{-2.85}$ & -6.50$^{+3.47}_{-3.60}$ & -7.86$^{+3.20}_{-2.81}$ & -6.72$^{+3.67}_{-3.45}$ & -9.87$^{+0.95}_{-1.36}$ & -10.45$^{+1.08}_{-1.01}$ & 0.05$^{+0.54}_{-0.62}$\\
    $1 \times \sigma_{\mathrm{white}}$ & 1.841$^{+0.024}_{-0.014}$ & 3119$^{+559}_{-313}$ & -2.24$^{+0.81}_{-0.98}$ & -8.48$^{+2.41}_{-2.16}$ & -7.21$^{+3.49}_{-3.05}$ & -7.63$^{+2.89}_{-2.66}$ & -7.12$^{+3.37}_{-3.20}$ & -10.16$^{+1.25}_{-1.14}$ & -9.63$^{+1.61}_{-1.52}$ & 0.17$^{+0.58}_{-0.98}$\\
    $3 \times \sigma_{\mathrm{white}}$ & 1.859$^{+0.044}_{-0.036}$ & 3459$^{+1078}_{-485}$ & -2.21$^{+1.38}_{-1.39}$ & -7.96$^{+2.53}_{-2.46}$ & -6.41$^{+3.37}_{-3.40}$ & -7.46$^{+3.27}_{-2.95}$ & -6.65$^{+3.03}_{-3.15}$ & -9.99$^{+1.28}_{-1.23}$ & -7.84$^{+1.76}_{-2.61}$ & -1.01$^{+1.23}_{-1.21}$\\
    $10 \times \sigma_{\mathrm{white}}$ & 1.897$^{+0.042}_{-0.041}$ & 4134$^{+1355}_{-898}$ & -2.45$^{+1.29}_{-1.49}$ & -7.84$^{+2.60}_{-2.51}$ & -5.84$^{+3.31}_{-3.77}$ & -7.14$^{+3.41}_{-2.98}$ & -6.61$^{+3.34}_{-3.08}$ & -9.83$^{+1.40}_{-1.33}$ & -8.50$^{+2.16}_{-2.16}$ & -0.94$^{+1.16}_{-1.18}$\\
    \noalign{\smallskip}
    \hline
    \end{tabular}
    \caption{Retrieved atmospheric parameters of WASP-178\,b from HST data and simulated CSST observations.}
    \label{tab:wasp-178b}
\end{table*}

However, for all three cases (b, c and d), where red noise dominates white noise, a systematic upward offset appears in the reconstructed spectrum. This offset is likely associated with the injected time-correlated trends, which can introduce structures partially degenerate with the transit signal. In particular, when the red-noise trend exhibits a dip around the transit phase, it can be partially absorbed into the transit depth during fitting and cannot be fully removed by GP detrending, leading to an upward bias in the reconstructed spectrum. We further verified this behavior with a simple test: when the injected red-noise trend instead shows a convex shape during transit, the resulting reconstructed spectrum exhibits an offset in the opposite direction.

As an additional non-GP detrending check, we repeated the light-curve fitting using a simple linear baseline model, which is substantially different from the GP used to generate the injected red noise. 
This test gives broadly consistent spectral morphology for case b, but shows larger biases in the stronger red-noise case, supporting the use of GP covariance modeling as our fiducial treatment for time-correlated noise.

In practice, the offsets can be mitigated through joint multi-band fitting or by detrending with white-light curves, in which correlated noise is modeled more consistently. The primary impact of this bias is on the absolute transit depth, leading to an overestimation of the planetary radius and atmospheric temperature in retrieval analyses.

To assess the impact of such systematics on atmospheric parameter derivations, the same retrievals were performed on the reconstructed spectrum. As shown in Fig.~\ref{fig:wasp-178b-compare}, the MP spectrum follows the overall theoretical trend but is offset upward by approximately 0.001 in transit depth. Due to the dominance of H$^-$ continuum opacity, the spectra in the GV and GI bands are largely featureless, and thus the retrieved abundances of species in Table~\ref{tab:wasp-178b} such as TiO and VO are not meaningfully constrained. In contrast, the SiO abundance remains better constrained in case (b), marginally detectable in case (c), and becomes poorly constrained with broad posterior distributions in case (d), reflecting the increasing impact of red noise on weak spectral features. However, both the planetary radius and the atmospheric temperature are overestimated, resulting in a global shift in the spectrum.

Overall, correlated noise can bias the absolute spectral level and degrade parameter constraints. In the idealized white-noise-only case, the simulated CSST observations yield smaller uncertainties than the corresponding HST data. Although CSST has a smaller aperture than HST (2.0 m versus 2.4 m), in the UV/optical wavelength range, the HST/ACS total efficiency is about 60\% of expected CSST total efficiency \citep{Zhan2021}. Under realistic ($1\sigma$) red-noise conditions, the uncertainties become comparable or slightly larger, with weaker abundance constraints. Although the precision of individual spectral points remains comparable to HST, the need to cover multiple bands reduces observational efficiency. Given that significant features for WASP-178\,b are mainly confined to the UV, prioritizing selected bands may provide a more efficient observing strategy.

\subsection{Robustness Against Random Noise Realizations}\label{sec:random}

\begin{figure}
    \centering
    \includegraphics[width=0.7\linewidth]{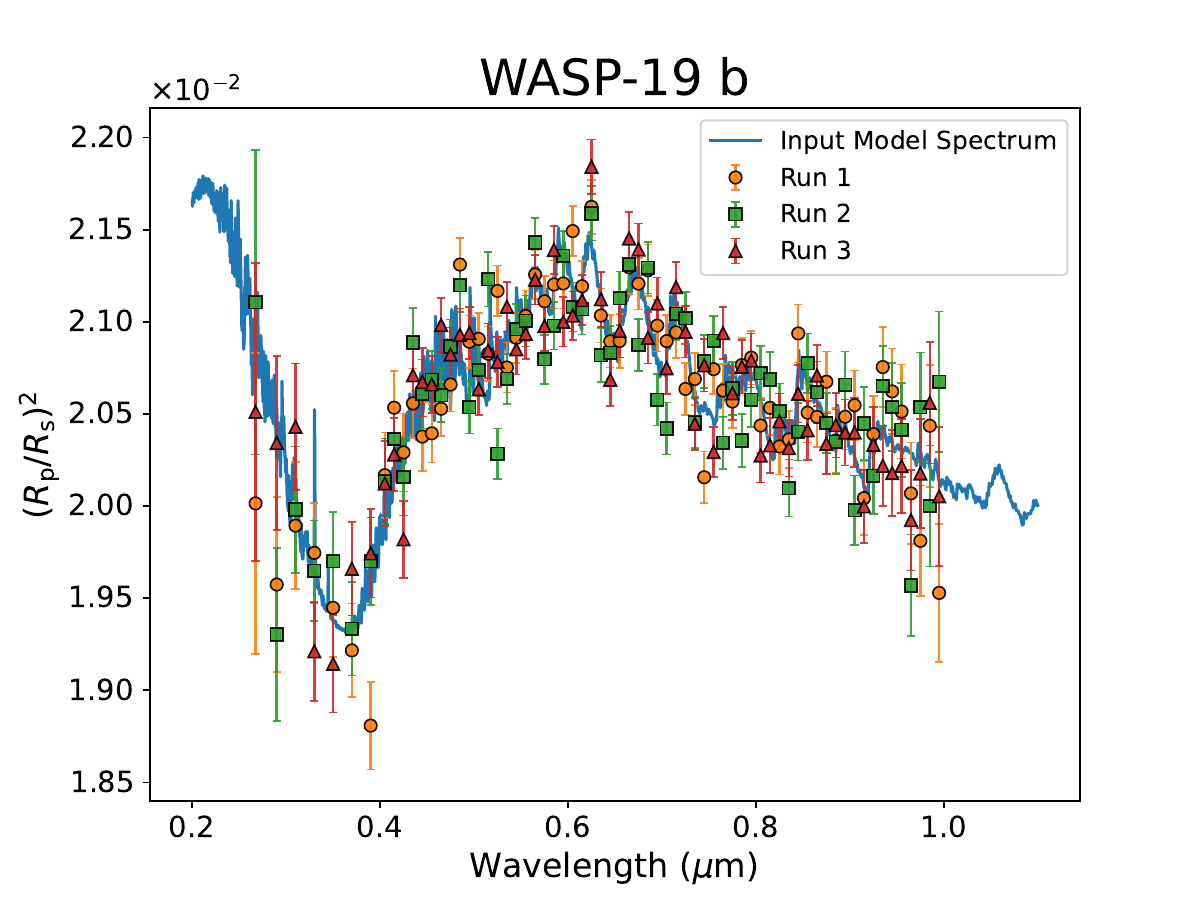}
    \caption{Simulated spectra of WASP-19\,b with three different random seeds.}
    \label{fig:WASP-19b-30min-test1}
\end{figure}

To assess the stability of our simulations against stochastic noise fluctuations, we repeated the spectral simulations for WASP-19\,b using different random seeds under the 30 min visibility scenario (Fig.~\ref{fig:WASP-19b-30min-test1}).

\begin{figure}
    \centering
    \includegraphics[width=1\linewidth]{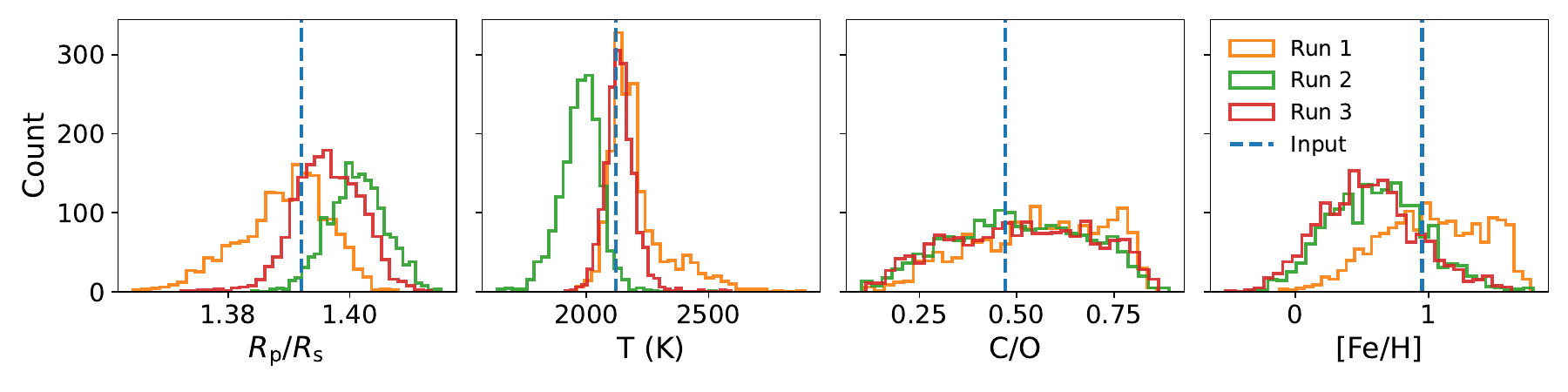}
    \caption{Retrieved atmospheric parameters of WASP-19\,b with three different random seeds, assuming chemical equilibrium.}
    \label{fig:WASP-19b-30min-test2}
\end{figure}

Although individual spectral points vary slightly between realizations, the overall spectral morphology remains consistent with the input theoretical model. The retrieved atmospheric parameters from equilibrium-chemistry fits in Fig.\ref{fig:WASP-19b-30min-test2} show MP values close to the theoretical inputs.

We performed two-sample Kolmogorov–Smirnov (K–S) tests to compare the three simulated spectra. The resulting p-values are 0.95 for run 1 vs. run 2 and run 1 vs. run 3, and 0.86 for run 2 vs. run 3. These values indicate that the spectra are statistically consistent with being drawn from the same distribution.

This test demonstrates that our simulation and retrieval framework is statistically robust and not sensitive to specific noise realizations.

\section{Conclusion} \label{sec:conclusion}

This work presents simulated transmission spectroscopic observations of seven gaseous exoplanets to evaluate the expected capability of CSST for atmospheric characterization of exoplanets. Using both equilibrium-chemistry and free-chemistry retrieval frameworks, we assessed the precision with which key atmospheric parameters can be constrained.

Our main conclusions are as follows:
\begin{enumerate}
\item Future CSST observations could characterize the atmospheres of hot and warm gas giants, place reliable constraints on atmospheric temperature, metallicity, and, in favorable cases, the C/O ratio. In the white-noise limit, the achievable precision can exceed that of HST data, while under realistic noise conditions it is comparable.
\item Multiple chemical species, including Na, K, H$_2$O, and SiO, are detectable within CSST’s wavelength coverage, particularly for planets with moderate to high TSM.
\item Even with time-correlated noise, the overall spectral morphology and major opacity sources remain recoverable, although biases in the absolute spectral level can weaken abundance constraints, yielding precision comparable to or slightly worse than HST depending on the noise level and observing strategy.
\item Planets with higher TSM values (TSM $\gtrsim 80$) yield quite robust atmospheric constraints, while lower-TSM targets benefit significantly from multiple transit observations.
\end{enumerate}

The expected UV sensitivity of CSST could provide strong, unique leverage on continuum slopes and short-wavelength absorbers, enabling tight constraints on atmospheric structure and key opacity sources. Although the spectral resolution and wavelength coverage limit the number of detectable molecules, the retrieved abundances of dominant species can provide the most important insights into bulk atmospheric composition and planetary formation history. In this context, future CSST observations could serve as a valuable complement to current JWST measurements by providing access to UV and blue-optical diagnostics that are not accessible to JWST. Together, such multi-wavelength coverage could help break degeneracies and improve constraints on atmospheric properties. This will be particularly valuable before the advent of future UV-capable space missions such as Tianlin and Habitable Worlds Observatory (HWO).

\section*{Acknowledgments}

%\begin{acknowledgments}
This research is supported by the National Key R\&D Program of China (2025YFE0213204, 2025YFE0102100, 2024YFA1611802), the National Natural Science Foundation of China under grants 62127901, 12588202, and the National Astronomical Observatories of the Chinese Academy of Sciences No. E4TQ2101, and the Pre-research project on Civil Aerospace Technologies No. D010301 funded by the China National Space Administration (CNSA), by the China Manned Space Program with grant no. CMS-CSST-2025-A16. \\
This work is based on the scientific data-processing software system for the Chinese Space Station Telescope under the China Manned Space Project.
%\end{acknowledgments}

%% To help institutions obtain information on the effectiveness of their 
%% telescopes the AAS Journals has created a group of keywords for telescope 
%% facilities.
%
%% Following the acknowledgments section, use the following syntax and the
%% \facility{} or \facilities{} macros to list the keywords of facilities used 
%% in the research for the paper.  Each keyword is check against the master 
%% list during copy editing.  Individual instruments can be provided in 
%% parentheses, after the keyword, but they are not verified.

\vspace{5mm}
\facilities{CSST}

%% Similar to \facility{}, there is the optional \software command to allow 
%% authors a place to specify which programs were used during the creation of 
%% the manuscript. Authors should list each code and include either a
%% citation or url to the code inside ()s when available.

\software{\texttt{astropy} \citep{astropy2013,astropy2018,astropy2022},  
          \texttt{matplotlib} \citep{Hunter2007}, 
          \texttt{petitRADTRANS} \citep{Molliere2019},
          \texttt{batman} \citep{Kreidberg2015},\texttt{SLS\_1D\_SPEC},\texttt{emcee} \citep{Foreman-Mackey2013},\texttt{george} \citep{Ambikasaran2015}}

%% Appendix material should be preceded with a single \appendix command.
%% There should be a \section command for each appendix. Mark appendix
%% subsections with the same markup you use in the main body of the paper.

%% Each Appendix (indicated with \section) will be lettered A, B, C, etc.
%% The equation counter will reset when it encounters the \appendix
%% command and will number appendix equations (A1), (A2), etc. The
%% Figure and Table counter will not reset.

%% For this sample we use BibTeX plus aasjournals.bst to generate the
%% the bibliography. The sample631.bib file was populated from ADS. To
%% get the citations to show in the compiled file do the following:
%%
%% pdflatex sample631.tex
%% bibtext sample631
%% pdflatex sample631.tex
%% pdflatex sample631.tex

\bibliography{reference}{}
\bibliographystyle{aasjournal}

\clearpage
\appendix

\section{Effect of Shorter Orbital Visibility} \label{sec:app-a}

Since CSST will operate at an altitude of $\sim$400 km, lower than that of HST, the available observing time per orbit is expected to be shorter than the typical $\sim$45 min of HST. To evaluate the impact of reduced visibility, we performed additional simulations for WASP-52\,b and WASP-19\,b assuming per-orbit observing windows of 30 min and 20 min.

\renewcommand{\thefigure}{\thesection\arabic{figure}}
\setcounter{figure}{0}

\begin{figure}[htbp]
    \centering
    \includegraphics[width=1\textwidth]{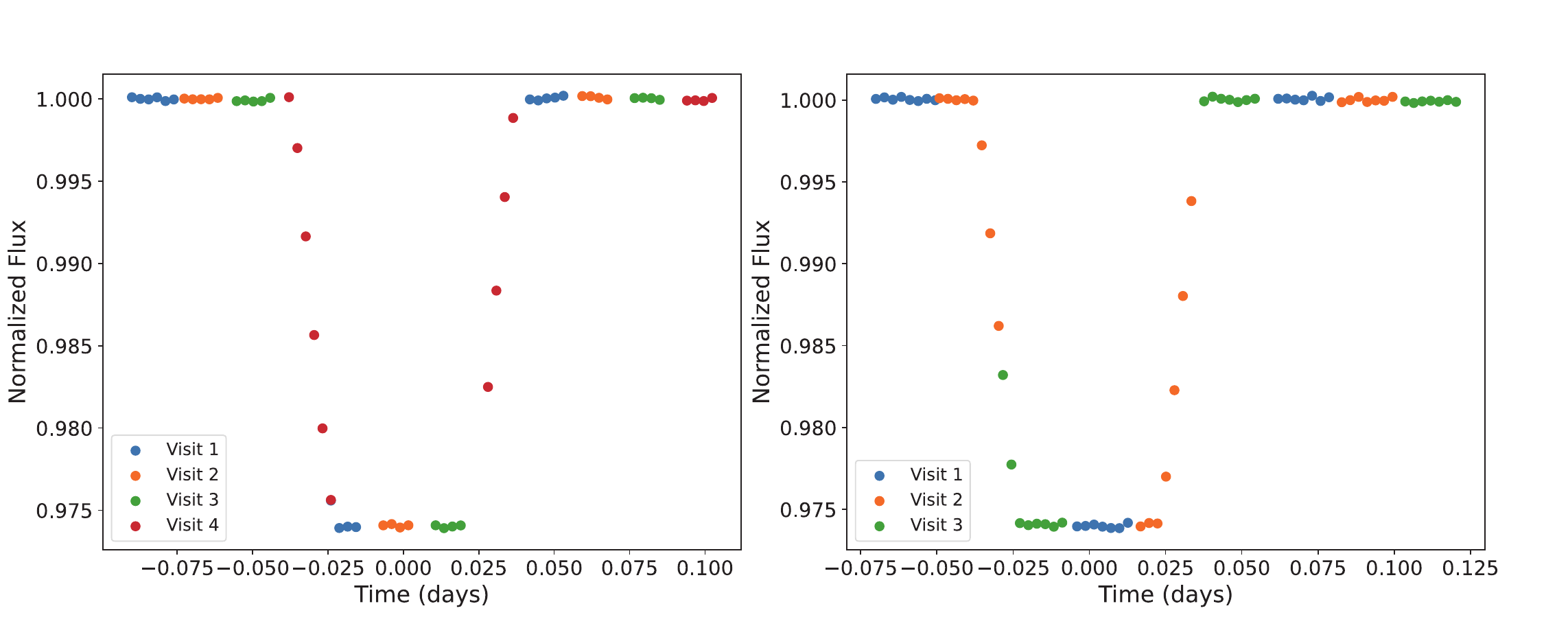}
    \caption{Phase-folded light curve of WASP-52\,b with 20 and 30 minutes observation window.}
    \label{fig:WASP-52b-20min-transit}
\end{figure}

For the 30 min case, three transit observations were combined, while for the 20 min case, four transits were stacked to ensure sufficient phase coverage (As shown in Fig.~\ref{fig:WASP-52b-20min-transit}). The resulting transmission spectra exhibit uncertainties comparable to those obtained under the 45 min visibility assumption, as the total accumulated exposure time is similar.

These tests indicate that reduced orbital visibility could be mitigated by increasing the number of orbits and repeated transit observations, at the expense of additional observing time and telescope resources, while maintaining comparable spectral precision.

\begin{figure}[htbp]
    \centering
    \gridline{\fig{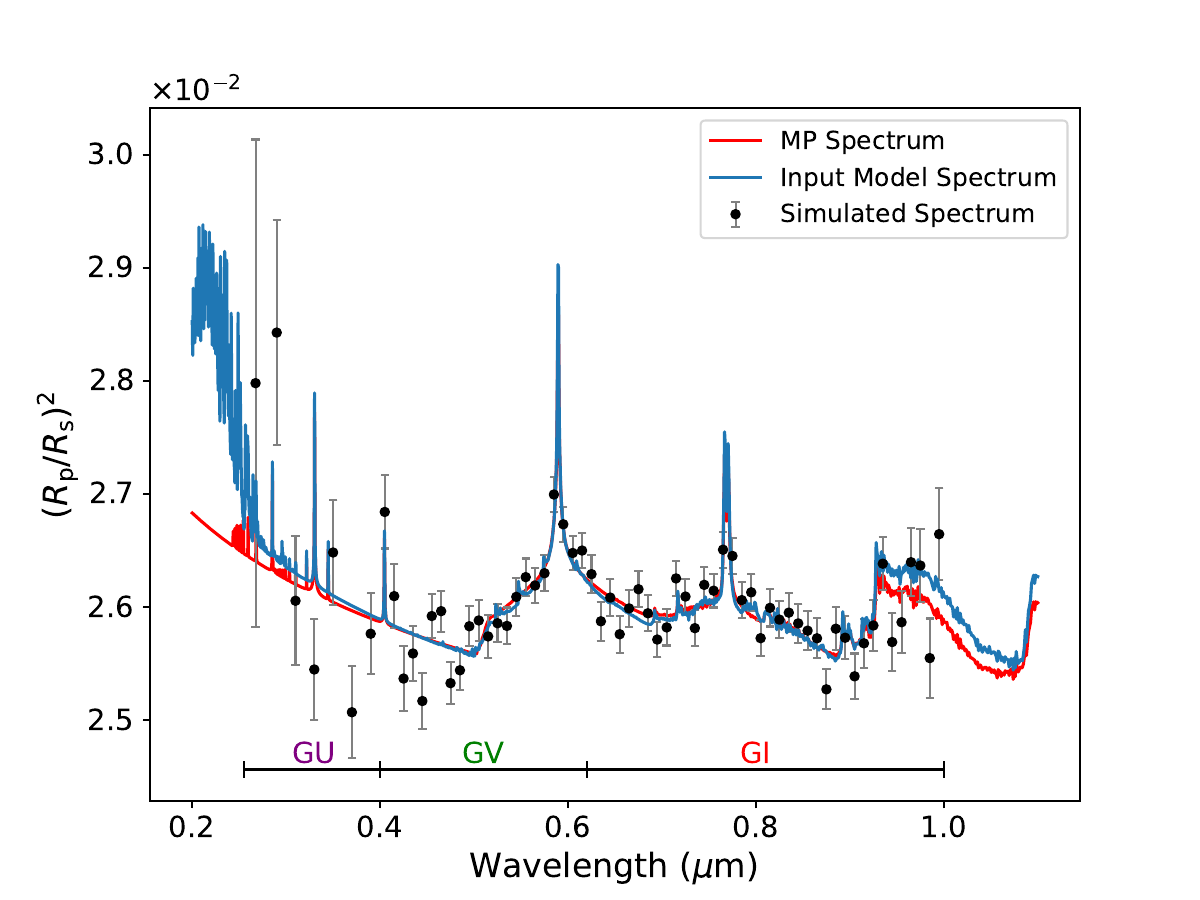}{0.5\linewidth}{(a) 20 minutes, 4 transits}
    \hspace{-0.05\textwidth}
    \fig{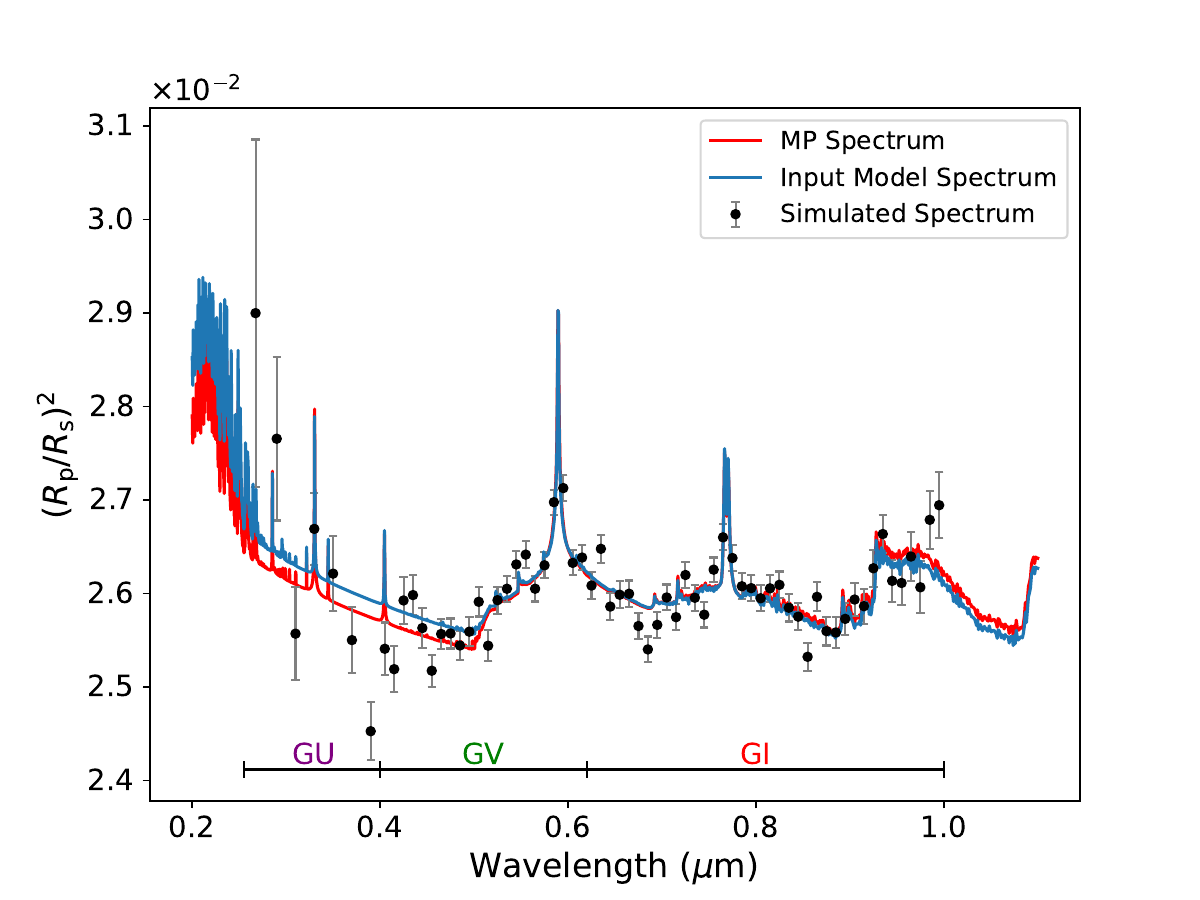}{0.5\linewidth}{(b) 30 minutes, 3 transits}}
    \caption{Equilibrium chemistry retrievals of WASP-52\,b with 20 and 30 minutes observation window.}
    \label{fig:WASP-52b-20min-eq}
\end{figure}

\begin{figure}[htbp]
    \centering
    \includegraphics[width=1\textwidth]{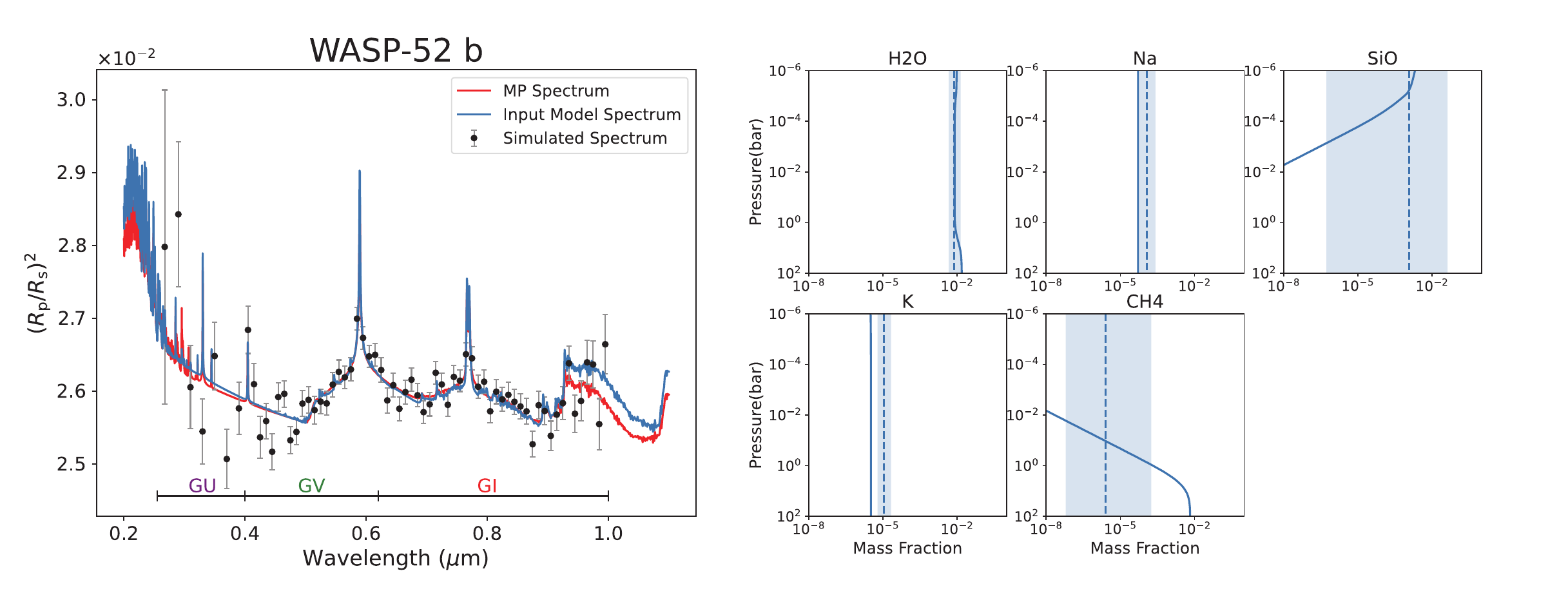}
    \caption{Free chemical retrievals of WASP-52\,b with 20 minutes observation window.}
    \label{fig:WASP-52b-20min-free}
\end{figure}

\begin{figure}[htbp]
    \centering
    \includegraphics[width=1\textwidth]{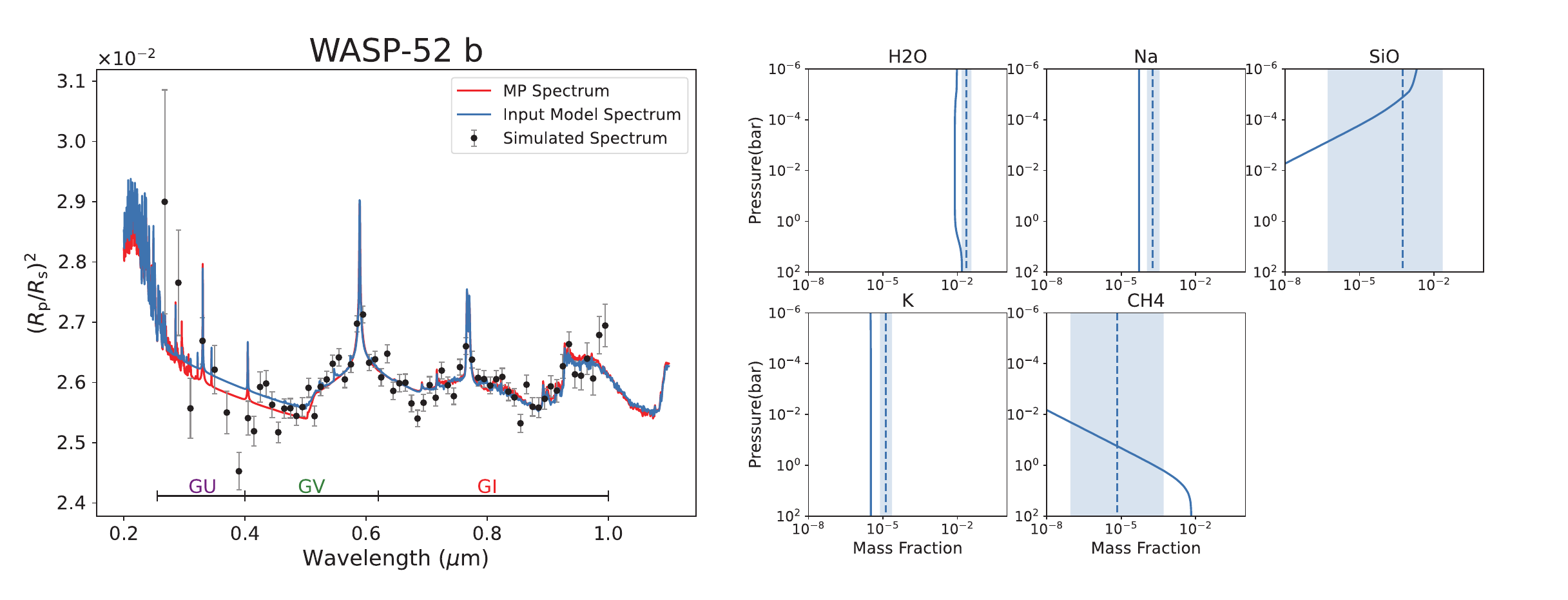}
    \caption{Same as Fig.\ref{fig:WASP-52b-20min-free}, but for 30 minutes observation window.}
    \label{fig:WASP-52b-30min-free}
\end{figure}

\begin{figure}[htbp]
    \centering
    \gridline{\fig{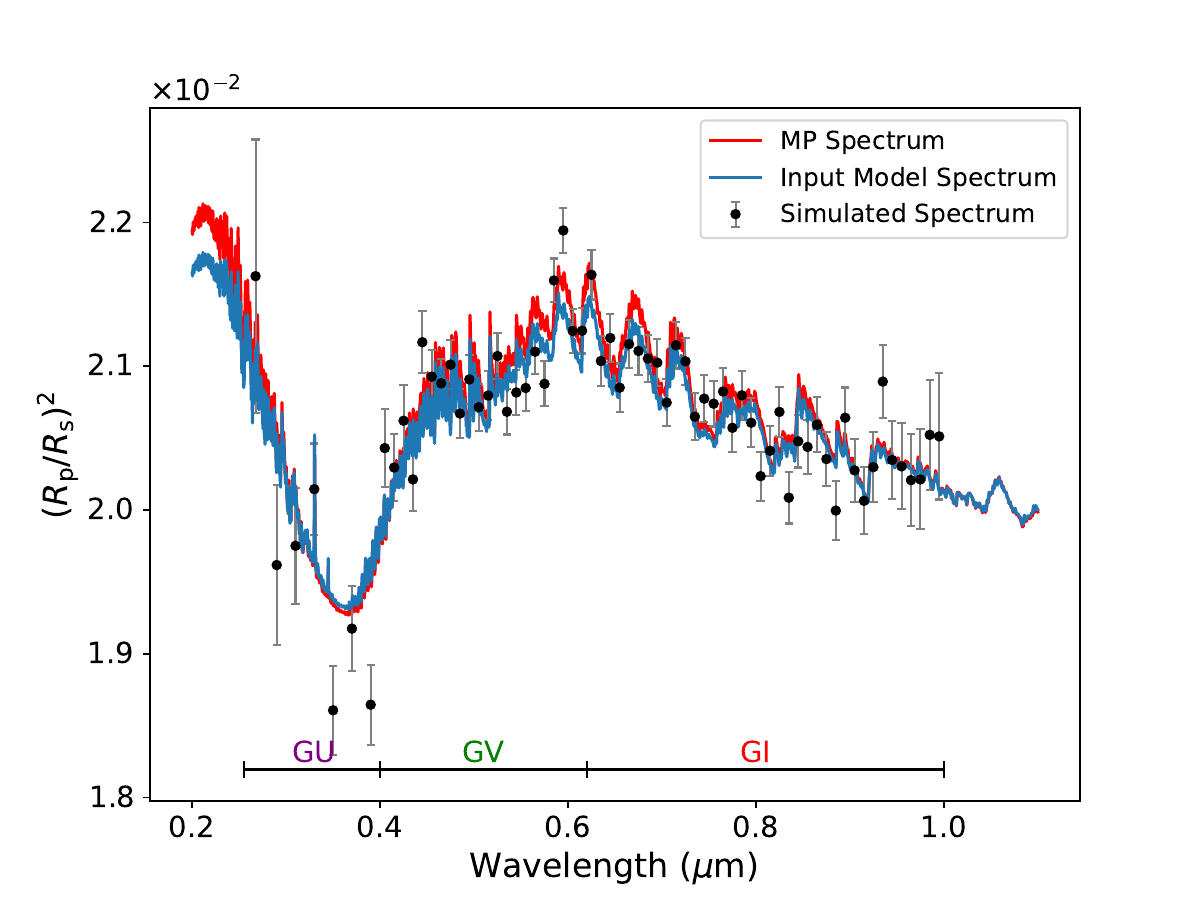}{0.5\linewidth}{(a) 20 minutes, 4 transits}
    \hspace{-0.05\textwidth}
    \fig{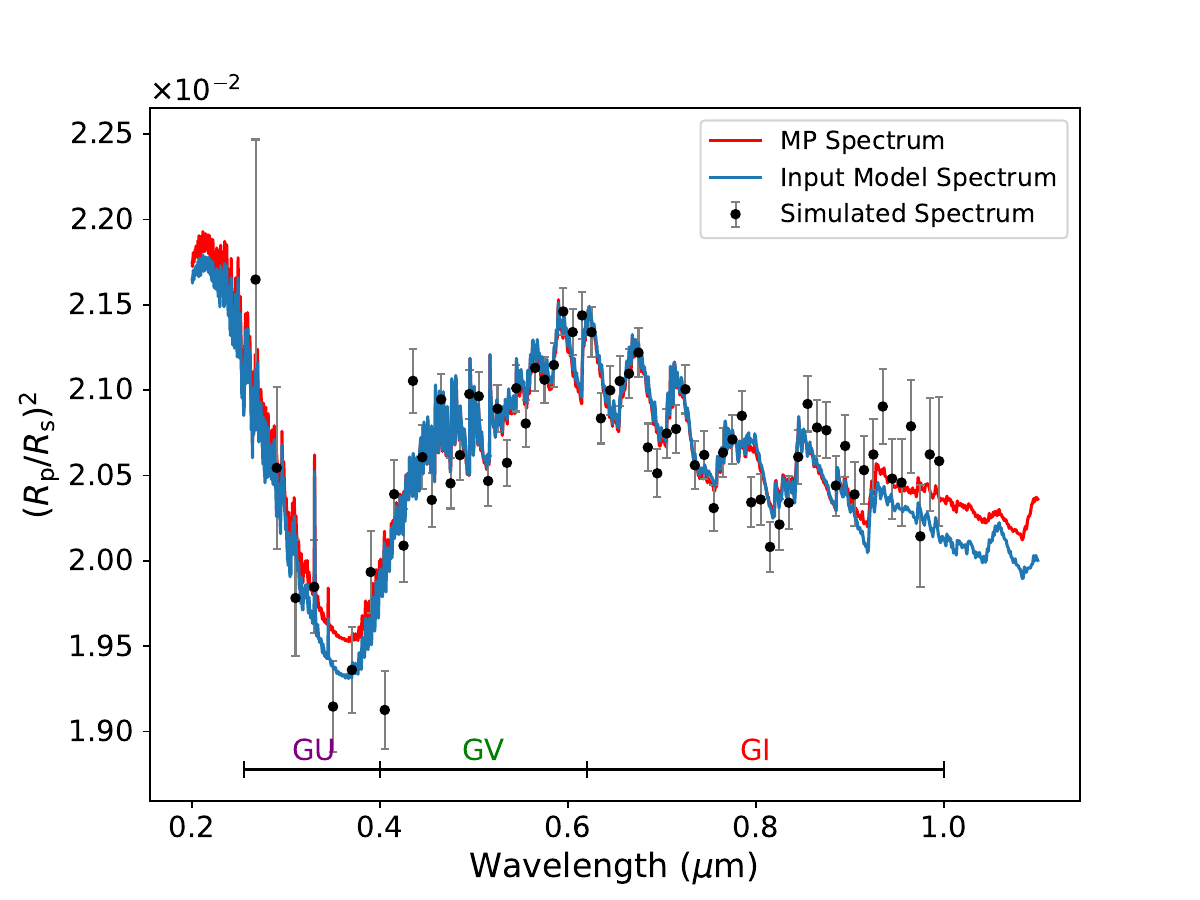}{0.5\linewidth}{(b) 30 minutes, 3 transits}}
    \caption{Same as Fig.\ref{fig:WASP-52b-20min-eq}, but for WASP-19\,b.}
    \label{fig:WASP-19b-20min-eq}
\end{figure}

\begin{figure}[htbp]
    \centering
    \includegraphics[width=1\textwidth]{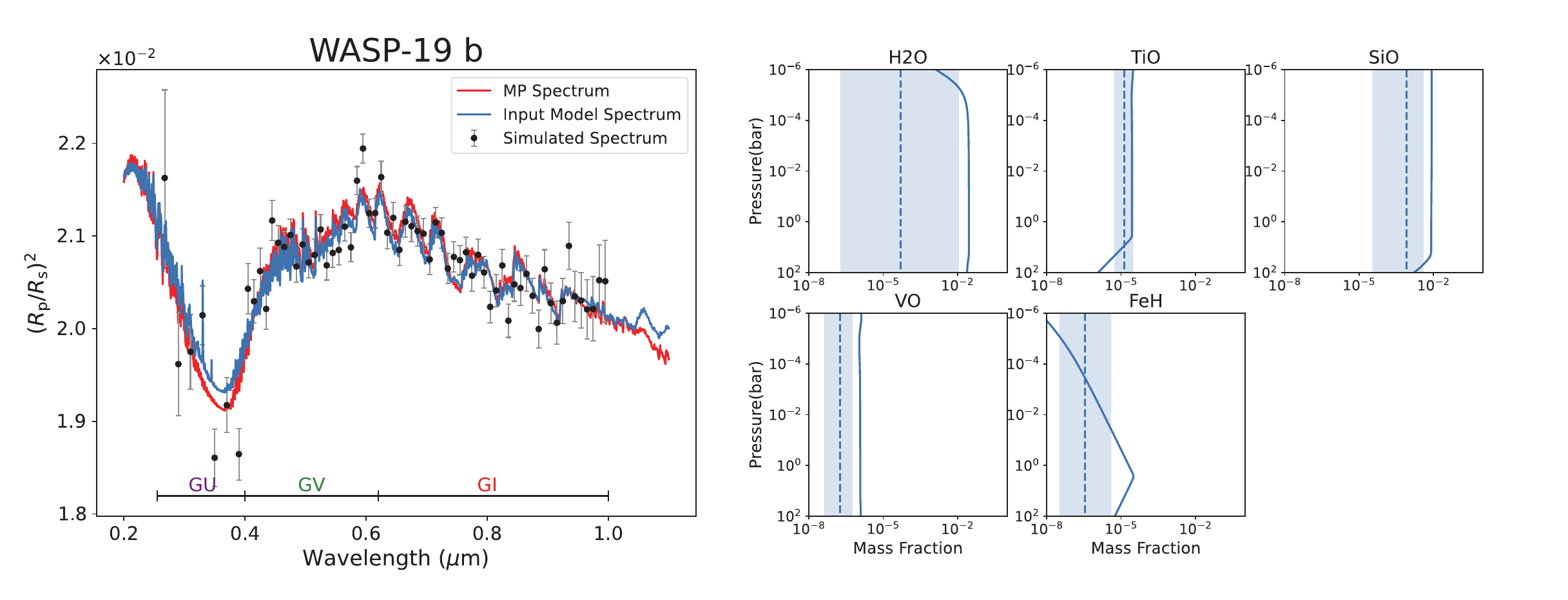}
    \caption{Same as Fig.\ref{fig:WASP-52b-20min-free}, but for WASP-19\,b and 20 minutes observation window.}
    \label{fig:WASP-19b-20min-free}
\end{figure}

\begin{figure}[htbp]
    \centering
    \includegraphics[width=1\textwidth]{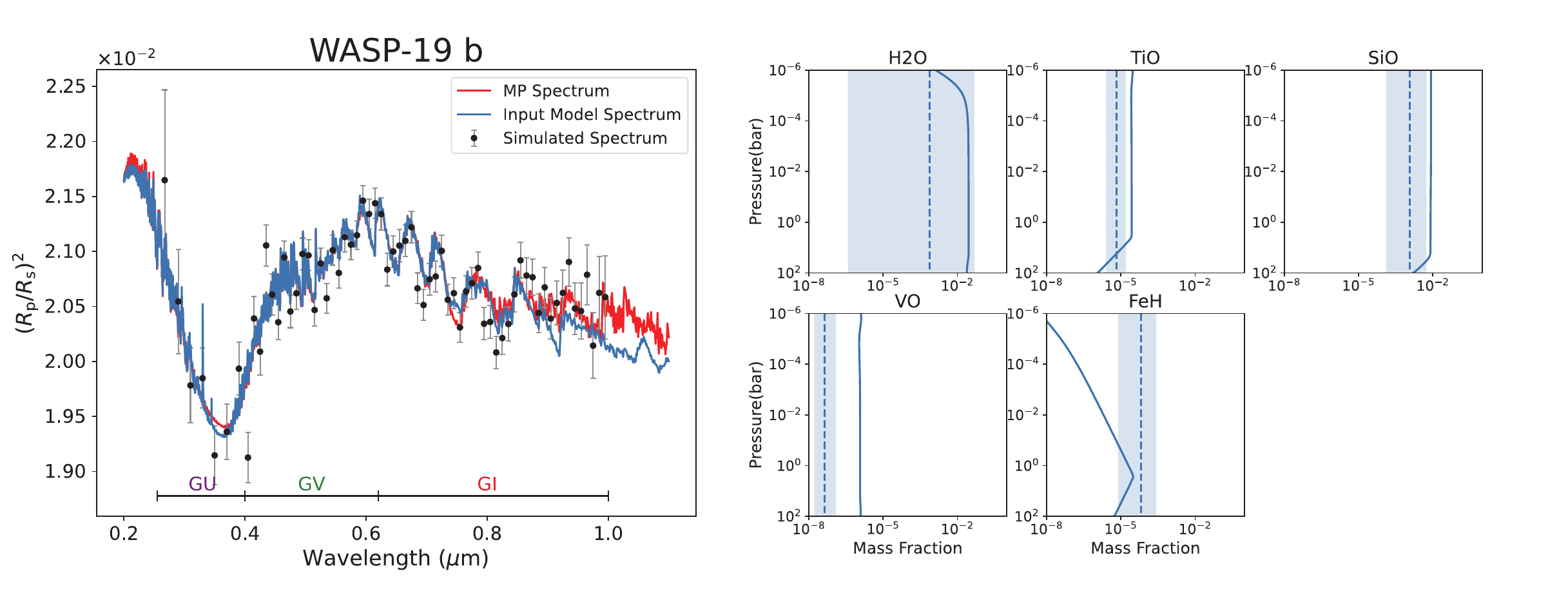}
    \caption{Same as Fig.\ref{fig:WASP-52b-20min-free}, but for WASP-19\,b and 30 minutes observation window.}
    \label{fig:WASP-19b-30min-free}
\end{figure}

\clearpage

%% This command is needed to show the entire author+affiliation list when
%% the collaboration and author truncation commands are used.  It has to
%% go at the end of the manuscript.
%\allauthors

%% Include this line if you are using the \added, \replaced, \deleted
%% commands to see a summary list of all changes at the end of the article.
%\listofchanges

\end{document}